\documentclass[a4paper,11pt]{article}
\pdfoutput=1 % if your are submitting a pdflatex (i.e. if you have
\usepackage{jheppub} % for details on the use of the package, please
\usepackage{etoolbox}
\makeatletter
    \patchcmd{\maketitle}{\@fpheader}{}{}{}
    \makeatother
    
\usepackage[T1]{fontenc} % if needed
\usepackage{subcaption}

\newcommand{\PRLsep}{\noindent\makebox[\linewidth]{\resizebox{0.5\linewidth}{1pt}{$\bullet$}}\smallskip}

\title{All Holographic Four-Point Functions in All Maximally Supersymmetric CFTs}

\author[]{Luis F. Alday$^{a}$,}
\affiliation[]{$^{a}$Mathematical Institute, University of Oxford,
Andrew Wiles Building, Radcliffe Observatory Quarter, Woodstock Road, Oxford, OX2 6GG, U.K.}
\author[]{Xinan Zhou$^{b}$}
\affiliation[]{$^{b}$Princeton Center for Theoretical Science, Princeton University, Princeton, NJ 08544, USA }

\emailAdd{alday@maths.ox.ac.uk}
\emailAdd{xinanz@princeton.edu}

\abstract{We present a constructive derivation of holographic four-point correlators of arbitrary half-BPS operators for all maximally supersymmetric conformal field theories in $d>2$. This includes holographic correlators in 3d ${\cal N}=8$ ABJM theories, 4d ${\cal N}=4$ SYM theory and the 6d ${\cal N}=(2,0)$ theory, dual to tree-level amplitudes in 11D supergravity on $AdS_4 \times S^7$, 10D supergravity on $AdS_5 \times S^5$ and 11D supergravity on $AdS_7 \times S^4$, respectively. We introduce the concept of Maximally R-symmetry Violating (MRV) amplitude, which corresponds to a special configuration in the R-symmetry space. In this limit the amplitude drastically simplifies, but at the same time the entire polar part of the full amplitude can be recovered from this limit. Furthermore, for a specific choice of the polar part, contact terms can be shown to be absent, by using the superconformal Ward identities and the flat space limit. }

\begin{document}
\maketitle
\flushbottom
\section{Introduction}
The AdS/CFT duality remains to this day the best  tool to  study physics at strong coupling analytically. Yet twenty two years since its discovery \cite{Maldacena:1997re,Gubser:1998bc,Witten:1998qj}, we are still on our way to harnessing the full computational power of this correspondence. The duality is simplest to study when there is a maximal amount of superconformal symmetry ({\it i.e.}, sixteen supercharges). This leads to three possibilities\footnote{We will focus on boundary theories in $d>2$. Correlators in SCFT$_2$ have special features which do not generalize to higher spacetime dimensions.}: 
\begin{itemize}
\item M-theory on $AdS_4\times S^7$ dual to the 3d $\mathcal{N}=8$ Aharony-Bergman-Jafferis-Maldacena (ABJM) theory \cite{Aharony:2008ug}, with superconformal group $OSp(8|4)$;
\item IIB string theory on $AdS_5\times S^5$ dual to 4d $\mathcal{N}=4$ Super Yang-Mills theory, with superconformal group $PSU(2,2|4)$;
\item M-theory on $AdS_7\times S^4$  dual to the 6d $\mathcal{N}=(2,0)$ theory, with superconformal group $OSp(8^*|4)$.
\end{itemize}
Furthermore, the bulk description becomes most tractable in the classical regime of supergravity, which occurs when the central charge approaches infinity (and with string length tending to zero in the $AdS_5\times S^5$ case). However, even in this regime and for these cases, much more should be computed than what's currently available in the literature. The most basic observables of the AdS/CFT correspondence are the correlation functions of local one-half BPS operators. Two-point and three-point functions are trivial because they are fully determined by superconformal symmetry\footnote{The three-point functions of one-half BPS operators are determined up to overall coefficients by conformal symmetry alone. The three-point coefficients are independent of marginal deformations (absent in the M-theory cases) thanks to supersymmetry.}. Only starting at four-points we begin to probe the nontrivial dynamics due to strong coupling. However, computing these correlators using holography, even at tree level, is long known to be notoriously difficult. The correlators in principle can be computed from a diagrammatic expansion in AdS, by following a standard procedure similar to the one for flat space QFTs. But one needs to extract all the relevant vertices from a complicated Kaluza-Klein (KK) reduction on the internal manifold $S^{\mathtt{d}-1}$, and there is an explosion of diagrams when considering operators dual to higher KK modes. These difficulties render the algorithm near impossible after a few low-lying cases \cite{DHoker:1999pj,Arutyunov:2000py,Arutyunov:2002ff,Arutyunov:2002fh,Arutyunov:2003ae,Berdichevsky:2007xd,Uruchurtu:2008kp,Uruchurtu:2011wh}. 

This situation becomes even more embarrassing when contrasted with the beautiful progress made for flat space scattering amplitudes (see, {\it e.g.}, \cite{Elvang:2015rqa,nima} for textbook presentations). Holographic correlators can be naturally identified with the on-shell scattering amplitudes in Anti de Sitter space. It would be truly surprising that no interesting structures can be found in holographic correlators. Motivated by this analogy with  flat space amplitudes and benefitting from developments in the  conformal bootstrap, a brand new method was proposed in \cite{Rastelli:2016nze,Rastelli:2017udc}, using the Mellin space representation \cite{Mack:2009mi,Penedones:2010ue}. By solely using  symmetry principles and consistency conditions, \cite{Rastelli:2016nze,Rastelli:2017udc} obtained a stunningly simple formula  for all tree-level four-point Mellin amplitudes for $AdS_5\times S^5$, as the solution to an algebraic bootstrap problem. This method eschewed the explicit details of the effective Lagrangian, and avoided all diagrams altogether. The general formula was later confirmed in a large number of explicit examples \cite{Arutyunov:2017dti,Arutyunov:2018neq,Arutyunov:2018tvn}\footnote{It also agrees with a previous conjecture for equal-weight correlators by Dolan, Nirschl and Osborn \cite{Dolan:2006ec}.},  and also provided essential data for studying correlators at one-loop \cite{Aharony:2016dwx,Alday:2017xua,Aprile:2017bgs,Aprile:2017xsp,Alday:2017vkk,Aprile:2017qoy,Aprile:2018efk,Caron-Huot:2018kta,Alday:2018pdi,Alday:2018kkw,Aprile:2019rep,Alday:2019nin}. The remarkable success of the method on $AdS_5\times S^5$ was partially replicated on $AdS_7\times S^4$, where the bootstrap problem was set up in \cite{Rastelli:2017ymc,Zhou:2017zaw}. Unfortunately, the problem was too difficult to be solved in general and only partial solutions for small weights were obtained \cite{Rastelli:2017ymc,Zhou:2017zaw}. Moreover, the same approach for $AdS_5$ and $AdS_7$ was not applicable to $AdS_4\times S^7$, due to a difference in the superconformal structure of correlators. A complementary method was subsequently introduced in \cite{Zhou:2017zaw}, which introduced superconformal Ward identities in Mellin space, and can be applied to any spacetime dimensions. However, this method also becomes cumbersome for more general correlators, and only the $AdS_4\times S^7$ stress tensor four-point function was explicitly written down \cite{Zhou:2017zaw}.

In this paper we solve {\it all} three theories with one single method, by borrowing new ideas from the flat space amplitudes. We constructively derive all tree-level four-point functions with arbitrary conformal dimensions, in all backgrounds with maximal superconformal symmetry. The result for $AdS_7\times S^4$ was already reported in an earlier publication \cite{Alday:2020lbp}, while for $AdS_5\times S^5$ our results give a proof of \cite{Rastelli:2016nze,Rastelli:2017udc}. Our method is based on the crucial observation that the supergravity Mellin amplitudes admit special limits of R-symmetry configuration, where drastic simplifications occur. In these special configurations, we align the R-symmetry polarizations of two operators. We will call such configurations Maximally R-symmetry Violating (MRV), in analogy with Maximally Helicity Violating (MHV) in the flat space parlance. The MRV amplitudes display two striking features: there are no singularities in the $u$ Mandelstam-Mellin variable; and the amplitude develops a factor of two zeroes in $u$. The first property follows from the fact that the supergravity field exchanges in the u-channel are suppressed by the special choice of R-symmetry configuration. The second property is the manifestation of the decoupling of low-lying unprotected long operators in this limit. Both features are universal for correlators from all theories with sixteen supercharges. However, at the level of individual Witten diagrams the consequence of the u-channel zeroes is highly nontrivial.  Quite remarkably,  all the field exchanges in each individual super multiplet must conspire in order to produce the zeroes. Imposing the presence of zeroes fixes the contribution of all component fields in the multiplet up to an overall constant, which can be easily computed by using the bulk cubic couplings of scalar fields. The zeroes in the MRV limit also forbids adding further contact terms, and allows us to write the MRV amplitudes in terms of a sum of simple multiplet exchange amplitudes. However, the study of the MRV amplitudes serves a greater purpose. From the MRV limit, we can use R-symmetry to restore the full dependence on the R-symmetry cross ratios, in the multiplet exchange amplitudes.  This determines the full correlators up to an addition of possible contact terms.  Note that contact terms can be mixed into the exchange amplitudes under field redefinitions. However, they are not arbitrary once a choice for the exchange amplitudes is made, and are uniquely fixed by requiring that the correlators satisfy superconformal Ward identities. We will provide a prescription to recover exchange amplitudes from the MRV limit such that no explicit contact terms are present. Using this procedure, we construct all tree-level four-point functions in $AdS_4\times S^7$, $AdS_5\times S^5$ and $AdS_7\times S^4$.

The rest of the paper is organized as follows. In Section \ref{Sec:2.1} we review the basic kinematics of four-point functions of one-half BPS operators. In Section \ref{Sec:2.2}, we review the traditional diagrammatic expansion method, and various bootstrap methods. We study the properties of the MRV limit in Section \ref{Sec:3}, and present an efficient algorithm for constructing all MRV amplitudes. In Section \ref{Sec:4}, we show how to recover the full amplitude from the MRV limit, and present the general result for all four-point functions in the three maximally superconformal backgrounds. In Section \ref{Sec:5}, we address the absence of contact terms by studying superconformal Ward identities in Mellin space. We also study these Ward identities and Mellin amplitudes near the flat space limit. We conclude in Section \ref{Sec:6}, and outline a few future directions. Various technical details are relegated to the two appendices.

\section{Generalities}\label{Sec:2}
\subsection{Kinematics}\label{Sec:2.1}
We focus on the one-half BPS local operators in superconformal field theories which have sixteen supercharges. Such operators $\mathcal{O}_k^{I_1\ldots I_k}$ transform in the rank-$k$ symmetric traceless representation of an $SO(\mathtt{d})$ R-symmetry group, with $k=2,3\ldots$. They have protected conformal dimension $\Delta_k=\epsilon k$, where $\epsilon$ is related to the spacetime dimension $d$ via $\epsilon=\frac{d-2}{2}$. It is convenient to keep track of the R-symmetry indices by contracting them with null vectors
\begin{equation}
\mathcal{O}_k(x,t)=\mathcal{O}_k^{I_1,\ldots,I_k}(x)t_{I_1} \ldots t_{I_k}\;,\quad t\cdot t=0\;.
\end{equation}
The four-point functions are denoted by 
\begin{equation}
G_{k_1k_2k_3k_4}(x_i,t_i)=\langle \mathcal{O}_{k_1}\mathcal{O}_{k_2}\mathcal{O}_{k_3}\mathcal{O}_{k_4}\rangle\;,
\end{equation}
and are functions of both the spacetime coordinates $x_i$ and internal coordinates $t_i$. We will often leave the $k_i$ dependence in $G_{k_1k_2k_3k_4}(x_i,t_i)$ implicit to avoid overloading the notation. We can assume, without loss of generality, that the weights $k_i$ are ordered as $k_1\leq k_2\leq k_3\leq k_4$. Then we need to further distinguish two possibilities 
\begin{equation}\label{twocases}
k_1+k_4\geq k_2+k_3 \;\; \text{(case I)}\;,\quad\quad\;\; k_1+k_4< k_2+k_3 \;\; \text{(case II)}\;.
\end{equation}
We can extract a kinematic factor 
\begin{equation}\label{GandcalG}
G(x_i,t_i)=\prod_{i<j}\left(\frac{t_{ij}}{x_{ij}^{2\epsilon}}\right)^{\gamma^0_{ij}}\left(\frac{t_{12}t_{34}}{x_{12}^{2\epsilon}x_{34}^{2\epsilon}}\right)^{\mathcal{E}}\mathcal{G}(U,V;\sigma,\tau)\;,
\end{equation}
such that the correlators can be written as a function of the cross ratios
\begin{equation}
U=\frac{x_{12}^2x_{34}^2}{x_{13}^2x_{24}^2}\;,\;\;\quad V=\frac{x_{14}^2x_{23}^2}{x_{13}^2x_{24}^2}\;,\;\;\quad \sigma=\frac{t_{13}t_{24}}{t_{12}t_{34}}\;,\;\;\quad\tau=\frac{t_{14}t_{23}}{t_{12}t_{34}}\;.
\end{equation}
Here $x_{ij}=x_i-x_j$, $t_{ij}=t_i\cdot t_j$, and $\mathcal{E}$ is the {\it extremality}
\begin{equation}
\mathcal{E}=\frac{k_1+k_2+k_3-k_4}{2} \;\;\; \text{(case I)}\;,\;\;\quad\quad \mathcal{E}=k_1 \;\;\; \text{(case II)}\;.
\end{equation}
The exponents are given by
\begin{eqnarray}
&&\gamma_{12}^0=\gamma_{13}^0=0\;,\;\; \gamma_{34}^0=\frac{\kappa_s}{2}\;,\; \; \gamma_{24}^0=\frac{\kappa_u}{2}\;,\\
\nonumber && \gamma_{14}^0=\frac{\kappa_t}{2}\,,\;\; \gamma_{23}^0=0\,,\;\text{(I)}\,,\;\;\;\gamma_{14}^0=0\,,\;\; \gamma_{23}^0=\frac{\kappa_t}{2}\,,\;\text{(II)}
\end{eqnarray}
where
\begin{equation}
\kappa_s\equiv|k_3+k_4-k_1-k_2|\;,\;\; \kappa_t\equiv|k_1+k_4-k_2-k_3|\;,\;\; \kappa_u\equiv|k_2+k_4-k_1-k_3|\;.
\end{equation}
Since $t_i$ can only appear in $G(x_i,t_i)$ as polynomials of $t_{ij}$, and $G(x_i,\lambda_i t_i)=\big(\prod_i\lambda_i^{k_i}\big)\, G(x_i,t_i)$ under rescaling, it is clear from (\ref{GandcalG}) that $\mathcal{G}(U,V;\sigma,\tau)$ is a polynomial in $\sigma$ and $\tau$ of degree $\mathcal{E}$. Writing $G(x_i,t_i)$ as in (\ref{GandcalG}) exploits only the bosonic part of the superconformal group. Fermionic generators imply further constraints, known as the superconformal Ward identities. It is useful to introduce the following change of variables
\begin{equation}
U=z\bar{z}\;,\quad\quad V=(1-z)(1-\bar{z})\;,\quad\quad \sigma=\alpha\bar{\alpha}\;,\quad\quad \tau=(1-\alpha)(1-\bar{\alpha})\;.
\end{equation}
The superconformal Ward identity reads \cite{Dolan:2004mu}
\begin{equation}\label{scfWardid}
(z\partial_z-\epsilon \alpha\partial_\alpha)\mathcal{G}(z,\bar{z};\alpha,\bar{\alpha})\big|_{\alpha=1/z}=0\;. 
\end{equation}
Because $\mathcal{G}(z,\bar{z};\alpha,\bar{\alpha})$ is symmetric under $z\leftrightarrow\bar{z}$ and $\alpha\leftrightarrow\bar{\alpha}$, three more identities follow from the above identity by replacing $z$ with $\bar{z}$, and $\alpha$ with $\bar{\alpha}$. 

\subsection{Methods for computing holographic correlators}\label{Sec:2.2}
\subsubsection{The traditional method: diagrammatic expansion}
The traditional recipe to calculate holographic correlators follows from a standard diagrammatic expansion in AdS. More precisely, one obtains the effective action on $AdS_{d+1}$, by performing a Kaluza-Klein reduction of the $D$ dimensional supergravity theory on $S^{D-d-1}$. For tree-level four-point functions, the relevant information to be extracted from the effective action are the cubic and quartic vertices. One then uses these vertices to write down all the possible exchange and contact Witten diagrams, and the four-point correlator is given by the sum
\begin{equation}\label{GasGexchGcon}
\mathcal{G}_{\rm tree}= \mathcal{G}^{(s)}_{\rm exch}+\mathcal{G}^{(t)}_{\rm exch}+\mathcal{G}^{(u)}_{\rm exch}+\mathcal{G}_{\rm con}\;.
\end{equation}
Here the number of exchanged fields in a specific four-point function is always finite. They are dictated by two selection rules on the cubic couplings. The first is an R-symmetry selection rule, which says that the R-symmetry representation carried by the exchanged fields (say in the s-channel) must appear in the common tensor product of the external representations ({\it i.e.},  the overlap of the tensor product of rank $k_1$, $k_2$ symmetric traceless representations, and that of $k_3$, $k_4$). The second is a cutoff on the conformal twist of the exchanged fields
\begin{equation}
\Delta-\ell<\epsilon\min{\{k_1+k_2,k_3+k_4\}}\;,
\end{equation}
which arises from the requirement that the effective action must remain finite. We organize the relevant exchanged fields into superconformal multiplets in the table below \cite{Kim:1985ez,Biran:1983iy,Castellani:1984vv,vanNieuwenhuizen:1984iz}, where the super primary scalar field $s_p$ is the bulk dual of the one-half BPS operator $\mathcal{O}_p$. The fields $A_{p,\mu}$ and $C_{p,\mu}$ are vector fields in $AdS$, and $A_{2,\mu}$ is the graviphoton field dual to the R-symmetry currents on the boundary. $\varphi_{p,\mu\nu}$ are the symmetric traceless spin-2 tensor fields, which include the graviton with $p=2$, dual to the stress tensor operator. $t_p$ and $r_p$ are scalar fields.
{\begin{center}
 \begin{tabular}{||c| c | c | c | c | c | c ||} 
 \hline
 field & $s_p$ & $A_{p,\mu}$ & $\varphi_{p,\mu\nu}$ &  $C_{p,\mu}$ & $t_p$ & $r_p$ \\ [0.5ex] 
 \hline\hline
$\ell$ & 0 & 1 & 2 & 1& 0 & 0\\ 
 \hline
$\Delta$ & $\epsilon p$ & $\epsilon p+1$ & $\epsilon p+2$ & $\epsilon p+3$ & $\epsilon p+4$ & $\epsilon p+2$ \\
 \hline
$d_1$ &  $p$ & $p-2$ & $p-2$ & $p-4$ & $p-4$ & $p-4$ \\ 
\hline $d_2$ &  $0$ & $2$  & $0$  & $2$ & $0$ & $4$  \\ [0.5ex] 
 \hline
\end{tabular}
\end{center}}
\noindent In the table, the quantum numbers $d_1$, $d_2$ are associated with the R-symmetry representation of the component fields, and appear in the Dynkin labels as 
\begin{equation}
SO(5):\;\; [d_1,d_2]\;,\quad\quad SU(4):\;\; [\tfrac{d_2}{2},d_1,\tfrac{d_2}{2}]\;,\quad\quad SO(8):\;\; [d_1,\tfrac{d_2}{2},0,0]\;.
\end{equation}
We can write the exchange contributions more explicitly as 
\begin{eqnarray}
\mathcal{G}^{(s)}_{\rm exch}&=&\sum_p \mathcal{V}^{(s)}_p\;,\label{GsVsp}\\
 \mathcal{V}^{(s)}_p&=&\lambda_s\, \mathcal{Y}_{\{p,0\}} W^{(s)}_{\epsilon p,0}+\lambda_A\, \mathcal{Y}_{\{p-2,2\}} W^{(s)}_{\epsilon p+1,1}+\lambda_{\varphi}\,\mathcal{Y}_{\{p-2,0\}} W^{(s)}_{\epsilon p+2,2}\label{Vsp}\\
\nonumber&&+\lambda_C\, \mathcal{Y}_{\{p-4,2\}} W^{(s)}_{\epsilon p+3,1}+ \lambda_t\, \mathcal{Y}_{\{p-4,0\}} W^{(s)}_{\epsilon p+4,0}+\lambda_r\, \mathcal{Y}_{\{p-4,4\}} W^{(s)}_{\epsilon p+2,0} 
\end{eqnarray}
where $\mathcal{V}^{(s)}_p$ is the contribution from the multiplet $p$. Here $W^{(s)}_{\Delta,\ell}$ are the standard exchange Witten diagrams in the s-channel with dimension $\Delta$ and spin $\ell$. $\mathcal{Y}_{\{d_1,d_2\}}$ are R-symmetry polynomials of $\sigma$ and $\tau$  (see Appendix \ref{App:A} for details), associated with the exchanged irreducible representation labelled by the R-symmetry quantum numbers $\{d_1,d_2\}$. Historically, such R-symmetry structures were obtained by gluing together three-point spherical harmonics. However, it is more convenient to obtain them by solving the two-particle quadratic R-symmetry Casimir equation \cite{Nirschl:2004pa}, making $\mathcal{Y}_{\{d_1,d_2\}}$ the compact analogues of conformal blocks. The coefficients $\lambda_{\rm field}$ in (\ref{Vsp}) are pure numbers, which can be fixed by using the explicit cubic vertices and appropriately taking into account the normalization of $\mathcal{Y}_{\{d_1,d_2\}}$. Finally, $\mathcal{G}_{\rm con}$ contains contact Witten diagrams up to four derivatives, and all possible R-symmetry structures. The simplest zero-derivative contact Witten diagram is denoted by the $\bar{D}$-function $\bar{D}_{\Delta_1\Delta_2\Delta_3\Delta_4}$ in the literature, and higher-derivative contact diagrams can be related to the zero-derivative ones by using differential recursion relations. The contact diagram contribution could in principle be computed when the quartic vertices are known.

Though clear physically, the traditional method suffers from several severe practical drawbacks. First of all, extracting the vertices, especially the quartic vertices, from the effective action is extremely hard. The general quartic vertices are only known for IIB supergravity on $AdS_5\times S^5$ \cite{Arutyunov:1999fb}, where their complicated expressions filled 15 pages. Second, as one increases the external dimensions (more precisely, the extremality $\mathcal{E}$), one is greeted by a proliferation of exchange Witten diagrams. Finally, the exchange Witten diagrams are only tractable in position space when the quantum numbers are fine tuned. When the spectrum satisfies the conditions 
\begin{equation}\label{trunccondi}
\Delta_1+\Delta_2-(\Delta-\ell)\in 2\,\mathbb{Z}_+\;,\quad \text{or} \quad \Delta_3+\Delta_4-(\Delta-\ell)\in 2\,\mathbb{Z}_+\;,
\end{equation}
the exchange Witten diagrams can be written as a finite sum of contact diagrams \cite{DHoker:1999mqo}. This is the case for $AdS_5\times S^5$ and $AdS_7\times S^4$. However, the conditions are not satisfied by the $AdS_4\times S^7$ background. These practical difficulties make it clear that this brute force approach is extremely cumbersome at best, and unlikely to yield any general result unless powerful underlying organizing principles can be identified.

\subsubsection{Bootstrap methods}
In recent years, a number of powerful bootstrap methods \cite{Rastelli:2016nze,Rastelli:2017udc,Rastelli:2017ymc,Zhou:2017zaw,Zhou:2018ofp,Rastelli:2019gtj} have been developed to efficiently compute holographic correlators, which have superseded the traditional method. These bootstrap methods exploit symmetries and self-consistency conditions, and fix the correlators by making no reference to the explicit details of the effective Lagrangian. Below we give an overview for these methods, and discuss their respective strengths and limitations.  

\vspace{0.5cm}
\noindent {\bf The position space method.} A first improvement of the traditional algorithm was made in \cite{Rastelli:2016nze,Rastelli:2017udc}, and was termed the {\it position space method}. The idea is to leave $\lambda_{\rm field}$ in (\ref{Vsp}) as unfixed parameters, and parameterize the most general contact contribution $\mathcal{G}_{\rm con}$ with unknown coefficients. In models where the truncation conditions (\ref{trunccondi}) are satisfied, one can write the exchange Witten diagrams in terms of a finite number of $\bar{D}$-functions. Furthermore, the $\bar{D}$-functions can be uniquely decomposed as
\begin{equation}\label{posibasis}
R_\Phi(z,\bar{z}) \Phi(U,V)+R_{\log U}(z,\bar{z}) \log U+R_{\log V}(z,\bar{z}) \log V+R_1(z,\bar{z}) 
\end{equation}
where $\Phi(U,V)$ is the scalar box diagram in four dimensions, and the coefficient functions $R_X(z,\bar{z})$ are rational functions of $z$ and $\bar{z}$. One then imposes the superconformal Ward identities (\ref{scfWardid}), which can be cast into the same form (\ref{posibasis}) by using differential recursion relations of $\Phi(U,V)$. The superconformal Ward identities uniquely fix all the unknown coefficients in the ansatz, up to an overall rescaling factor. This method has the advantage of being very concrete, and sidesteps the need for obtaining the complicated vertices. On the other hand, the method is applied on a case by case basis, and runs out of steam for higher weight external operators. The position space method can be applied to supergravity theories on $AdS_5\times S^5$ \cite{Rastelli:2016nze,Rastelli:2017udc}, $AdS_7\times S^4$ \cite{Rastelli:2017ymc} and $AdS_3\times S^3\times K3$ \cite{Rastelli:2019gtj}\footnote{The $AdS_3\times S^3\times K3$ background preserves only eight supercharges. As a consequence, while contact diagrams are fixed by exchange coefficients, there are unfixed cubic couplings which require additional input.} backgrounds.\footnote{The position space method is not limited to supergravity fluctuations. In \cite{Drukker:2020swu} it was shown that the method also produces defect four-point functions of Wilson loops in 4d $\mathcal{N}=4$ SYM and surface defects in the 6d (2,0) theory, in  regimes where they are  world-volume fluctuations of probe strings and M2 branes in AdS.  } However, it is not applicable to 11D supergravity on $AdS_4\times S^7$ where the exchange Witten diagrams do not truncate. Finally, the expressions of holographic correlators in position space are usually  highly complicated, and beg for a more transparent representation which we now introduce below. 

\vspace{0.2cm}

\PRLsep

\vspace{0.5cm}
\noindent \textit{Intermezzo: Mellin space.} A useful tool for holographic correlators is the Mellin representation formalism \cite{Mack:2009mi,Penedones:2010ue}. This formalism was exploited in the methods below and later will also be the language of this paper. In the Mellin representation 
\begin{eqnarray}\label{inverseMellin}
\mathcal{G}_{\rm tree}&=&\int_{-i\infty}^{i\infty} \frac{dsdt}{(4\pi i)^2} U^{\frac{s}{2}-a_s}V^{\frac{t}{2}-a_t}\mathcal{M}(s,t;\sigma,\tau)\,\Gamma_{\{k_i\}}\;,\\
\nonumber \Gamma_{\{k_i\}}&=&\Gamma[\tfrac{\epsilon(k_1+k_2)-s}{2}]\Gamma[\tfrac{\epsilon(k_3+k_4)-s}{2}]\Gamma[\tfrac{\epsilon(k_1+k_4)-t}{2}]\Gamma[\tfrac{\epsilon(k_2+k_3)-t}{2}]\Gamma[\tfrac{\epsilon(k_1+k_3)-u}{2}]\Gamma[\tfrac{\epsilon(k_2+k_4)-u}{2}]
\end{eqnarray}
where $a_s=\frac{\epsilon}{2}(k_1+k_2)-\epsilon\mathcal{E}$, $a_t=\frac{\epsilon}{2}\min\{k_1+k_4,k_2+k_3\}$ and $s+t+u=\epsilon \sum_{i=1}^4 k_i\equiv \epsilon\Sigma$, the analytic structure of the holographic correlators becomes particularly clear. The Mellin amplitudes of exchange Witten diagrams are a sum over simple poles
\begin{equation}\label{Mellinexchange}
\mathcal{M}^{(s)}_{\Delta,\ell}(s,t)=\sum_{m=0}^\infty \frac{\mathcal{Q}_{m,\ell}(t,u)}{s-\Delta+\ell-2m}
\end{equation}
where $\mathcal{Q}_{m,\ell}(t,u)$ are degree-$\ell$ polynomials in $t$ and $u$. The residues $\mathcal{Q}_{m,\ell}(t,u)$ vanish for $m\geq m_0$ when the conditions (\ref{trunccondi}) are satisfied 
\begin{equation}\label{trunccondi}
\Delta_1+\Delta_2=\Delta-\ell+2m_0\;,\quad \text{or} \quad \Delta_3+\Delta_4=\Delta-\ell+2m_0\;,
\end{equation}
truncating the infinite series into a finite sum, in order to be consistent with the large $N$ expansion \cite{Rastelli:2017udc}. On the other hand, contact diagrams with $2L$ derivatives have Mellin amplitudes which are polynomials in the Mandelstam variables of degree $L$. Note that in the literature, it is also conventional to write (\ref{Mellinexchange}) as 
\begin{equation}
\mathcal{M}^{(s)}_{\Delta,\ell}(s,t)=\sum_{m=0}^\infty \frac{\widetilde{\mathcal{Q}}_{m,\ell}(t)}{s-\Delta+\ell-2m}+\mathcal{P}_{\ell-1}(s,t)
\end{equation}
where one Mandelstam variable is eliminated from $\mathcal{Q}_{m,\ell}(t,u)$, and $\mathcal{P}_{\ell-1}(s,t)$ is a degree-$(\ell-1)$ polynomial. In (\ref{Mellinexchange}) we have absorbed the regular terms into the numerator. This is related to the fact that exchange Witten diagrams are not uniquely defined. We can add to them any contact terms with degree $\ell-1$, which corresponds to choosing different on-shell equivalent cubic couplings. 
\vspace{0.2cm}

\PRLsep

\vspace{0.5cm}
\noindent {\bf The Mellin algebraic bootstrap method.} A more elegant method was formulated in \cite{Rastelli:2016nze,Rastelli:2017udc,Rastelli:2017ymc}, which rephrased the task of computing holographic four-point functions as solving an {\it algebraic bootstrap problem} in Mellin space. This method exploits the special structure of the correlators as dictated by the superconformal Ward identities
\begin{equation}
\mathcal{G}=\mathcal{G}_0+\mathcal{D}\circ \mathcal{H}
\end{equation}
where $\mathcal{G}_0$ is a protected part of the correlator that does not contribute to the Mellin amplitude. $\mathcal{D}$ is a differential operator determined by superconformal symmetry, and $\mathcal{H}$ is known as the {\it reduced correlator}. We can define a {\it reduced} Mellin amplitude $\widetilde{\mathcal{M}}$ from $\mathcal{H}$, and translate the differential operator $\mathcal{D}$ into a difference operator  $\widehat{\mathcal{D}}$ in Mellin space. Then we have
\begin{equation}
\mathcal{M}=\widehat{\mathcal{D}}\circ \widetilde{\mathcal{M}}\;,
\end{equation}
which implements the superconformal symmetry at the level of Mellin amplitudes. The bootstrap problem is formulated by further imposing Bose symmetry, analytic properties and flat space limit on the Mellin amplitude $\mathcal{M}$. Such algebraic bootstrap problems are highly constraining, and fix the correlators uniquely up to an overall constant. The bootstrap problem for $AdS_5\times S^5$ was fully solved in \cite{Rastelli:2016nze,Rastelli:2017udc} for arbitrary four-point functions, and led to an extremely compact answer. The merit of this approach is that one can treat all external dimensions on the same footing, and obtain the correlators without computing any diagrams. However, the analytic structure of the reduced amplitude $\widetilde{\mathcal{M}}$ is not as transparent as that of the full amplitude $\mathcal{M}$. This makes it sometimes difficult to find a general efficient ansatz for $\widetilde{\mathcal{M}}$, such as in $AdS_7\times S^4$, and the problem is solved only on a case by case basis \cite{Rastelli:2017ymc,Zhou:2017zaw}. Moreover, for $d=3$ the differential operator $\mathcal{D}$ is non-local, which makes it difficult to interpret in Mellin space.

\vspace{0.5cm}
\noindent {\bf Mellin superconformal Ward identities.} Complementary to the above Mellin algebraic bootstrap method, is another Mellin space technique that can be applied to any spacetime dimensions, first developed in \cite{Zhou:2017zaw}. This method can be viewed as the Mellin space parallel of the position space method. We can translate (\ref{GasGexchGcon}), (\ref{GsVsp}), (\ref{Vsp}) into
\begin{eqnarray}\label{Ssp}
&&\nonumber \mathcal{M}(s,t;\sigma,\tau)=\mathcal{M}^{(s)}_{\rm exch}+\mathcal{M}^{(t)}_{\rm exch}+\mathcal{M}^{(u)}_{\rm exch}+\mathcal{M}_{\rm con}\,,\\
\nonumber &&\mathcal{M}^{(s)}_{\rm exch}(s,t;\sigma,\tau)=\sum\nolimits_p \mathcal{S}^{(s)}_p(s,t;\sigma,\tau)\;,\\
 &&\mathcal{S}^{(s)}_p=\lambda_s\, \mathcal{Y}_{\{p,0\}} \mathcal{M}^{(s)}_{\epsilon p,0}+\lambda_A\, \mathcal{Y}_{\{p-2,2\}} \mathcal{M}^{(s)}_{\epsilon p+1,1} +\lambda_{\varphi}\,\mathcal{Y}_{\{p-2,0\}} \mathcal{M}^{(s)}_{\epsilon p+2,2}\\
\nonumber&&\quad\quad+\lambda_C\, \mathcal{Y}_{\{p-4,2\}} \mathcal{M}^{(s)}_{\epsilon p+3,1}+ \lambda_t\, \mathcal{Y}_{\{p-4,0\}} \mathcal{M}^{(s)}_{\epsilon p+4,0}+\lambda_r\, \mathcal{Y}_{\{p-4,4\}} \mathcal{M}^{(s)}_{\epsilon p+2,0}\;,
\end{eqnarray}
with unfixed $\lambda_{\rm field}$, and $\mathcal{M}_{\rm con}$ will be taken as an arbitrary degree-1 polynomial in $s$, $t$, and a degree-$\mathcal{E}$ polynomial in $\sigma$, $\tau$. Then we would like to impose the superconformal constraints from the superconformal Ward identities (\ref{scfWardid}).   This may appear difficult as only $U$ and $V$ appear in the definition (\ref{inverseMellin}), which is invariant under $z\leftrightarrow \bar{z}$. However, the superconformal Ward identity (\ref{scfWardid}) breaks the symmetry of $z$ and $\bar{z}$, and creates complicated branch cuts when rewritten in terms of $U$ and $V$. The observation of \cite{Zhou:2017zaw} is that we can take the sum of a holomorphic and an anti-holomorphic copy\footnote{We can also take the difference and then take out an overall factor of $z-\bar{z}$, as we will see in Section \ref{Sec:5}.}
\begin{equation}
(z\partial_z-\epsilon \alpha\partial_\alpha)\mathcal{G}(z,\bar{z};\alpha,\bar{\alpha})\big|_{\alpha=1/z}=0\;,\quad\;\; (\bar{z}\partial_{\bar{z}}-\epsilon \alpha\partial_\alpha)\mathcal{G}(z,\bar{z};\alpha,\bar{\alpha})\big|_{\alpha=1/\bar{z}}=0\;. 
\end{equation}
Then the coefficients can always be written in terms of polynomials in $U$ and $V$, which are easy to interpret as difference operators in Mellin space. These difference equations (graded by different powers of the spectator cross ratio $\bar{\alpha}$) constitute the {\it Mellin superconformal Ward identities}. Imposing these identities, one fixes all the coefficients in the ansatz, up to an overall constant. Note that in Mellin space exchange Witten diagrams are easy to write down for any spacetime dimension and conformal dimensions.  This greatly extends the range of applicability of this method.
Using this Mellin space technique, \cite{Zhou:2017zaw} obtained the first four-point correlator in $AdS_4\times S^7$ for the stress tensor multiplet, where all other methods had fallen short. On the other hand, the method suffers from the same shortcomings as the position space approach, in that it is difficult to go beyond individual correlators.

\vspace{0.5cm}
\noindent {\bf Other approaches.} There are other methods for computing holographic correlators by incorporating  bootstrap ideas. By using factorization and supersymmetric twistings, \cite{Goncalves:2019znr} computed the five-point function of one-half BPS operators in the stress tensor multiplet for IIB supergravity on $AdS_5\times S^5$. In $AdS_3$, there is also a method to construct four-point functions from the heavy-heavy-light-light limit, by using crossing and consistency with superconformal OPE \cite{Giusto:2018ovt,Giusto:2019pxc,Giusto:2020neo}. This approach complements to the bootstrap method in $AdS_3$ \cite{Rastelli:2019gtj}.

\section{Maximally R-symmetry Violating limit}\label{Sec:3}
\subsection{Properties of the MRV amplitudes}
While the full Mellin amplitudes appear rather complicated, there are special limits where the amplitudes simplify drastically and give a hint for their underlying organizing principles. One such limit is the Maximally R-symmetry Violating (MRV) limit, introduced in \cite{Alday:2020lbp}. In the ordering of $k_1\leq k_2\leq k_3\leq k_4$, the (u-channel) MRV limit is reached by setting $t_1=t_3$ for the auxiliary R-symmetry null vectors. This choice of null vectors means that in $G(x_i,t_i)$, $t_1$ cannot be contracted with $t_3$ and no $t_{13}$ can appear. In terms of the R-symmetry cross ratios, it corresponds to setting $\sigma=0$, $\tau=1$. We will denote the MRV amplitude as\footnote{When writing $\text{MRV}(s,t)$, we will implicitly assume it is in the u-channel if without further specifications.} 
\begin{equation}
{\rm MRV}(s,t)=\mathcal{M}(s,t;0,1)\;.
\end{equation}
Note that the MRV limit can also be defined in other channels: in s-channel it corresponds to $t_1=t_2$, and in t-channel it amounts to $t_2=t_3$ (case I) or $t_1=t_4$ (case II).\footnote{When considering the s-channel MRV limit, it is better to restore the $t_{ij}$ factor extracted in (\ref{GandcalG}). This avoids the confusion associated with $\sigma$, $\tau$ becoming infinite.} The three limits are related by Bose symmetry. Restricting to the MRV limit suppresses certain R-symmetry representations in that channel. For example, all the u-channel supergravity field exchanges are suppressed in the $\sigma=0$, $\tau=1$ limit because the R-symmetry polynomials all contain at least one power of $t_{13}$. This gives the first simplifying property of MRV amplitudes:

\vspace{0.2cm}

{\it The MRV amplitudes have no poles in the u-channel.}

\vspace{0.2cm}

\noindent Moreover, in such special R-symmetry configurations we are allowed to see the interesting phenomenon that the super primary is absent, whereas super descendants are present.\footnote{Of course, both are visible in a generic R-symmetry configuration.} In particular, let us consider the long super multiplet where the super primary is a double-trace operator of the schematic form $[:\mathcal{O}_{k_2}\partial^J\mathcal{O}_{k_4}:]_{\{d_1,d_2\}}$. In order for all super descendants (in particular the operator acted with $Q^4\bar{Q}^4$ which has maximal deviation in R-symmetry from the super primary) to have R-symmetry charges admissible in the tensor products of $\mathcal{O}_{k_1}\times \mathcal{O}_{k_3}$ and $\mathcal{O}_{k_2}\times \mathcal{O}_{k_4}$, the representation of the super primary must satisfy $d_1+d_2\leq 2\mathcal{E}+\kappa_t+\kappa_u-4$. This implies that in the MRV configuration, the R-symmetry polynomial associated with $\{d_1,d_2\}$ vanishes. Moreover, one can show that the only super descendant which contributes to this limit is  $Q^4\bar{Q}^4[:\mathcal{O}_{k_2}\partial^J\mathcal{O}_{k_4}:]_{\{d_1,d_2\}}$. Therefore we expect to see long operators (albeit not a super primary) in the u-channel MRV configuration with conformal twist of at least $\epsilon(k_2+k_4)+4$. This is reflected by the double pole at $u=\epsilon(k_2+k_4)+4$ in the $\Gamma_{\{k_i\}}$ factor in (\ref{inverseMellin}). Upon doing the inverse Mellin integral, we see a logarithmic singularity which is the hallmark of an unprotected long operator. On the other hand, this lower bound for logarithmic singularities cannot be further lowered, because  $\epsilon(k_2+k_4)$ is the minimal twist of the double-trace operators constructed from $\mathcal{O}_{k_2}$ and $\mathcal{O}_{k_4}$ for the super primaries of the long multiplets. This implies the second important property of the MRV amplitudes:

\vspace{0.2cm}

{\it The MRV amplitudes contain a factor of zeroes $(u-\epsilon k_2-\epsilon k_4)(u-\epsilon k_2-\epsilon k_4-2)$.}

\vspace{0.2cm}

\noindent These zeroes are precisely needed to cancel one of the double poles in $\Gamma_{\{k_i\}}$, such that no logarithmic singularities at these twists show up. 

\subsection{All MRV amplitudes}

These two properties of MRV amplitudes have profound consequences in understanding the structure of holographic correlators. In fact, the u-channel zeroes are satisfied by each individual super multiplet exchange in the s-channel (and separately, in the t-channel). This gives rise to an efficient way to fix the relative values of $\lambda_{\rm field}$ inside each multiplet. More precisely, 
we choose the contact terms in the exchange Witten diagrams (\ref{Mellinexchange}) by setting $t=\epsilon\Sigma-u-(\Delta-\ell)-2m$ in the numerators $\mathcal{Q}_{m,\ell}(t,u)$
\begin{equation}
P^{(s)}_{\Delta,\ell}(s,u)=\sum_{m=0}^\infty \frac{\mathcal{Q}_{m,\ell}(\epsilon\Sigma-u-(\Delta-\ell)-2m,u)}{s-\Delta+\ell-2m}\;.
\end{equation}
This choice corresponds to the so-called Polyakov-Regge blocks \cite{Mazac:2019shk,Sleight:2019ive} (see also \cite{Gopakumar:2016wkt,Gopakumar:2016cpb,Gopakumar:2018xqi} for related blocks), which have improved u-channel Regge behavior
\begin{equation}
P^{(s)}_{\Delta,\ell}(s,u)\to \frac{1}{s}\;,\quad \;\; s\to\infty\;,\;\; u\;\;\text{fixed}\;.
\end{equation}
For simplicity, we will focus on case I of (\ref{twocases}) in what follows, in addition to the ordering $k_1\leq k_2\leq k_3\leq k_4$. However,  in the next section when we assemble the ingredients into the final results and express them in terms of $\kappa_s$, $\kappa_t$, $\kappa_u$, the expressions will be valid for any ordering of $k_i$ thanks to Bose symmetry. The $SO(\mathtt{d})$  R-symmetry polynomials take the following values in the MRV limit
\begin{eqnarray}\label{YMRV}
\nonumber \mathcal{Y}_{\{p,0\}}^{\rm MRV}&\equiv&\mathcal{Y}_{\{p,0\}}(0,1)=\frac{(\tfrac{\kappa_t}{2}!)(\tfrac{p+k_2-k_1}{2}!)\Gamma[\tfrac{\mathtt{d}+p+k_2-k_1-2}{2}]\Gamma[\tfrac{\mathtt{d}+p+k_4-k_3-2}{2}]}{(\tfrac{\kappa_u}{2}!)(\tfrac{p+k_1-k_2}{2}!)\Gamma[\tfrac{\mathtt{d}-k_1-k_3+k_2+k_4-2}{2}]\Gamma[\tfrac{2p+\mathtt{d}-2}{2}]}\;,\\
\mathcal{Y}_{\{p,2\}}^{\rm MRV}&\equiv&\mathcal{Y}_{\{p,2\}}(0,1)=-\frac{(p+k_2-k_1+\mathtt{d}-2)(p+k_4-k_3+\mathtt{d}-2)}{(\mathtt{d}+p-2)(\mathtt{d}+2p-2)}\mathcal{Y}_{\{p,0\}}^{\rm MRV}\;,\\
\nonumber \mathcal{Y}_{\{p,4\}}^{\rm MRV}&\equiv&\mathcal{Y}_{\{p,4\}}(0,1)=\frac{4(\mathtt{d}-3)(\tfrac{p+k_2-k_1+\mathtt{d}-2}{2})_2(\tfrac{p+k_4-k_3+\mathtt{d}-2}{2})_2 }{(\mathtt{d}-2)(-(\mathtt{d}+p))_2(-\tfrac{(\mathtt{d}+2p)}{2})_2}\mathcal{Y}_{\{p,0\}}^{\rm MRV}\;.
\end{eqnarray}
Requiring the presence of the zeroes at every pole $s=\epsilon\, p+2m$ imposes strong constraints on $\lambda_{\rm field}$, and solves them in terms of $\lambda_s$
\begin{eqnarray}\label{sollambda}
\nonumber \lambda_A^{(p)}&=&\tfrac{\mathcal{Y}_{\{p,0\}}^{\rm MRV}}{\mathcal{Y}_{\{p-2,2\}}^{\rm MRV}}\tfrac{\epsilon  (k_1-k_2+p) (k_3-k_4+p)}{2 p (p \epsilon +2)}\lambda_s^{(p)} \;,\\
\nonumber \lambda_{\varphi}^{(p)}&=&\tfrac{\mathcal{Y}_{\{p,0\}}^{\rm MRV}}{\mathcal{Y}_{\{p-2,0\}}^{\rm MRV}}\tfrac{\epsilon ^2 (k_1-k_2+p) (k_3-k_4+p) (k_1 \epsilon -k_2 \epsilon +p \epsilon +2) (k_3 \epsilon -k_4 \epsilon +p \epsilon +2)}{16 (p \epsilon +1) (p \epsilon +2)^2 (p \epsilon +3)}\lambda_s^{(p)}\;,\\
\nonumber\lambda_C^{(p)}&=&\tfrac{\mathcal{Y}_{\{p,0\}}^{\rm MRV}}{\mathcal{Y}_{\{p-4,2\}}^{\rm MRV}}\tfrac{\epsilon ^3 (k_1-k_2+p-2) (k_1-k_2+p) (k_3-k_4+p-2) (k_3-k_4+p) (\epsilon  (k_1-k_2+p)+2) (\epsilon  (k_3-k_4+p)+2)}{32 (p-2) ((p-1) \epsilon +1) ((p-1) \epsilon +2) (p \epsilon +2)^2 (p \epsilon +3)}\lambda_s^{(p)}\;,\\
\nonumber \lambda_r^{(p)}&=&\tfrac{\mathcal{Y}_{\{p,0\}}^{\rm MRV}}{\mathcal{Y}_{\{p-4,0\}}^{\rm MRV}}\tfrac{\epsilon ^2 (\epsilon +2) (k_1-k_2+p-2) (k_1-k_2+p) (k_3-k_4+p-2) (k_3-k_4+p)}{8 (p-2) (p-1) (\epsilon +1) ((p-1) \epsilon +2) (p \epsilon +2)}\lambda_s^{(p)}\;,\\
\nonumber \lambda_t^{(p)}&=&\tfrac{\mathcal{Y}_{\{p,0\}}^{\rm MRV}}{\mathcal{Y}_{\{p-4,4\}}^{\rm MRV}}\tfrac{\epsilon ^4 (k_1-k_2+p-2) (k_1-k_2+p) (k_3-k_4+p-2) (k_3-k_4+p)}{256 ((p-2) \epsilon +1) ((p-2) \epsilon +2) ((p-1) \epsilon +1) }\\
&\times& \tfrac{ (\epsilon  (k_1-k_2+p-2)+2) (\epsilon  (k_1-k_2+p)+2) (\epsilon  (k_3-k_4+p-2)+2) (\epsilon  (k_3-k_4+p)+2)}{((p-1) \epsilon +2)^2 ((p-1) \epsilon +3) (p \epsilon +2) (p \epsilon +3)}\lambda_s^{(p)}\;.
\end{eqnarray}
Here we have added a superscript to the coefficients $\lambda^{(p)}_{\rm field}$ to emphasize that they belong to the $p$-th multiplet. 
Inserting the solutions into $\mathcal{S}^{(s)}_p$ in (\ref{Ssp}) leads to a great simplification. We obtain the following contribution from each super multiplet to the MRV limit
\begin{eqnarray}
\nonumber \mathcal{S}^{(s)}_p(s,t;0,1)&=& \sum_{m=0}^\infty\frac{4 \lambda_s^{(p)} (p \epsilon +1) (p \epsilon -\epsilon +1)}{(k_1-k_2-p) (k_4-k_3+p) (k_1 \epsilon -k_2 \epsilon -p \epsilon -2) (k_3 \epsilon -k_4 \epsilon -p \epsilon -2)}\\
&\times&\frac{(u-\epsilon k_2-\epsilon k_4)(u-\epsilon k_2-\epsilon k_4-2)}{(p+1)_{-2} (m+p \epsilon -\epsilon )_2}\bigg(\frac{f_{m,0}\big|_{\Delta_E=\epsilon p}}{s-\epsilon p-2m}\bigg)
\end{eqnarray}
where the u-channel zeroes are factored out, leaving just a sum over simple poles with constant residues. The terms in the brackets are just the scalar exchange Mellin amplitude at each simple pole, with $f_{m,\ell_E}$ defined in Appendix \ref{App:B}. Notice that the MRV amplitude for each multiplet does {\it not} depend on the R-symmetry group $SO(\mathtt{d})$.

To write down the full MRV amplitude we just need to sum over all multiplets, which is restricted to be finite by the selection rules
\begin{equation}\label{prange}
p-\max\{|k_1-k_2|,|k_3-k_4|\}=2\,,4\,,\ldots 2\mathcal{E}-2\;.
\end{equation}
The strength of the contribution from each multiplet, captured by $\lambda_s^{(p)}$, can be determined from the three-point functions of the super primaries 
\begin{equation}
\langle \mathcal{O}_{k_1}(x_1,t_1) \mathcal{O}_{k_2}(x_2,t_2) \mathcal{O}_{k_3}(x_3,t_3)\rangle=C_{k_1k_2k_3}^{(\epsilon)}(\alpha_1,\alpha_2,\alpha_3)\frac{t_{12}^{\alpha_3}t_{13}^{\alpha_2}t_{23}^{\alpha_1}}{x_{12}^{2\epsilon \alpha_3}x_{13}^{2\epsilon \alpha_2}x_{23}^{2\epsilon \alpha_1}}
\end{equation}
where 
\begin{equation}
\alpha_1=\frac{1}{2}(k_2+k_3-k_1)\;,\quad \alpha_2=\frac{1}{2}(k_1+k_3-k_2)\;,\quad \alpha_3=\frac{1}{2}(k_1+k_2-k_3)\;.
\end{equation}
The three-point coefficients read \cite{Lee:1998bxa,Corrado:1999pi,Bastianelli:1999en}\footnote{Here we use $n$  to collectively denote the numbers of M2, D3 or M5 branes, while in the literature it is more conventional to use $N$ for the number of D3 branes.}\footnote{The extremal three-point functions ({\it i.e.}, $k_1+k_2=k_3$, {\it etc}) are a bit subtle. The finiteness of the bulk effective action requires these correlators to be zero. On the other hand, these three-point functions are non-vanishing in the field theory, and are given by the above formulae. This puzzle is solved by realizing that the supergravity states correspond to a mixture of single-trace and double-trace BPS operators, with mixing coefficients fixed precisely by the vanishing of extremal three-point functions \cite{Arutyunov:1999en,Arutyunov:2000ima}. This subtlety, however, does not affect our discussion, because all the poles in the Mellin amplitude are sub-extremal and are agnostic about the subtlety. In position space, mixing can affect the four-point functions by adding certain rational functions formed by product of two-point and three-point functions. However, the mixing effect cannot be detected by the Mellin amplitudes because the rational terms have zero Mellin amplitudes.}
\begin{eqnarray}
C^{(\frac{1}{2})}_{k_1k_2k_3}&=&\frac{\pi}{n^{\frac{3}{4}}}\frac{2^{-\alpha-\frac{1}{4}}}{\Gamma[\frac{\alpha}{2}+1]}\prod_{i=1}^3\frac{\sqrt{\Gamma[k_i+2]}}{\Gamma[\frac{\alpha_i+1}{2}]}\;,\\
C^{(1)}_{k_1k_2k_3}&=&\frac{\sqrt{k_1k_2k_3}}{n}\;,\\
C^{(2)}_{k_1k_2k_3}&=&\frac{2^{2\alpha-2}}{(\pi n)^{\frac{3}{2}}}\Gamma[\alpha]\prod_{i=1}^3 \frac{\Gamma[\alpha_i+\frac{1}{2}]}{\sqrt{\Gamma[2k_i-1]}}\;,
\end{eqnarray}
where $\alpha=\alpha_1+\alpha_2+\alpha_3$. $\lambda_s^{(p)}$ is given in terms of $C_{k_1k_2k_3}^{(\epsilon)}$ by 
\begin{equation}\label{lambdas}
\lambda_s^{(p)}=\left(\frac{(\tfrac{p+k_1-k_2}{2}!)(\tfrac{p+k_4-k_3}{2}!)}{p!(\tfrac{k_1+k_4-k_2-k_3}{2}!)}\right)C_{k_1k_2p}^{(\epsilon)}C_{k_3k_4p}^{(\epsilon)}
\end{equation}
where the number in the brackets is a gluing factor for the R-symmetry due to the fact that we have normalized the R-symmetry polynomials to have unit coefficients for $\sigma^{\mathcal{E}}$. The MRV amplitudes are then simply given by 
\begin{equation}
{\rm MRV}(s,t)={\rm MRV}^{(s)}(s,t)+{\rm MRV}^{(t)}(s,t)
\end{equation}
where 
\begin{equation}
{\rm MRV}^{(s)}(s,t)=\sum_p \mathcal{S}^{(s)}_p(s,t;0,1)\;,
\end{equation}
with the summation over $p$ inside the finite range (\ref{prange}), and ${\rm MRV}^{(t)}(s,t)$ is related to ${\rm MRV}^{(s)}(s,t)$ by Bose symmetry. Note that no additional contact terms are allowed in the MRV amplitudes. This follows from the simple fact that contact terms are at most linear in the Mandelstam variables, while the requisite zeroes are already quadratic. The absence of additional contact terms tells us something quite remarkable about the structure of supergravity theories in AdS: supersymmetry in  the MRV limit not only determines the relative cubic couplings of components within the same multiplet, but its implication reaches quartic couplings as well. It is also worth pointing out that the MRV amplitudes have an improved u-channel Regge behavior compared to a Witten diagram exchanging a spinning field and with generic choices of contact terms. The MRV amplitudes behave in the same way as the Polyakov-Regge blocks.

\section{All tree-level correlators from the MRV limit}\label{Sec:4}
\subsection{Full amplitudes from MRV amplitudes}
A lot more information can be extracted from the MRV limit. In fact, in constructing the MRV amplitudes we have determined all the polar part of the full Mellin amplitude. This follows from the fact that all R-symmetry polynomials (\ref{YMRV}) are non-vanishing in the MRV limit. We can therefore restore the full $\sigma$, $\tau$ dependence in (\ref{Ssp}) by using R-symmetry.\footnote{Note that the same happens for flat space graviton amplitudes in maximally super-symmetric gravity theories, since for four-point functions the full helicity dependence is contained in a universal prefactor ${\cal R}^4$.} More precisely, we can write down
\begin{eqnarray}\label{Stilde}
&&\widetilde{\mathcal{S}}^{(s)}_p=\lambda_s\, \mathcal{Y}_{\{p,0\}} P^{(s)}_{\epsilon p,0}+\lambda_A\, \mathcal{Y}_{\{p-2,2\}} P^{(s)}_{\epsilon p+1,1} +\lambda_{\varphi}\,\mathcal{Y}_{\{p-2,0\}} P^{(s)}_{\epsilon p+2,2}\\
\nonumber&&\quad\quad+\lambda_C\, \mathcal{Y}_{\{p-4,2\}} P^{(s)}_{\epsilon p+3,1}+ \lambda_t\, \mathcal{Y}_{\{p-4,0\}} P^{(s)}_{\epsilon p+4,0}+\lambda_r\, \mathcal{Y}_{\{p-4,4\}} P^{(s)}_{\epsilon p+2,0}\;,
\end{eqnarray}
where we have used the Polyakov-Regge blocks and it corresponds to a specific choice of contact terms. Various $\lambda^{(p)}_{\rm field}$ have been obtained in (\ref{sollambda}) and (\ref{lambdas}). It follows that $\widetilde{\mathcal{S}}^{(s)}_p$ gives the correct residues for any $\sigma$ and $\tau$.

However, note that the s-channel Polyakov-Regge blocks are not symmetric in $t$ and $u$. More precisely, the Bose symmetry in exchanging 1 and 2 is broken by the choice of the contact terms. This can be easily seen from the fact that the s-channel Polyakov-Regge blocks have improved Regge behavior in the u-channel, but not in the t-channel. To restore the s-channel Bose symmetry in the s-channel multiplet exchange, we give the following simple prescription \cite{Alday:2020lbp}. The amplitude $\widetilde{\mathcal{S}}^{(s)}_p$ takes the form of a sum over simple poles at $s=\epsilon p+2m$. For each term in the sum, the numerator contains a quadratic factor in $u$ of the form 
 \begin{equation}
u^2+\alpha(i,j;m,p)\, u+\beta(i,j;m,p)\;. 
\end{equation}
We can restore Bose symmetry, by eliminating $m$ from this factor from the relation 
\begin{equation}
t+u+\epsilon p+2m=\epsilon\Sigma
\end{equation}
where we have substituted the pole values of $s$ into the relation among the three Mandelstam variables. This gives a symmetric s-channel exchange, which we will denote as $\mathcal{S}^{(s)}_p$. Using the other generators of the Bose symmetry, we  can  similarly obtain $\mathcal{S}^{(t)}_p$ and $\mathcal{S}^{(u)}_p$. Note that our prescription is not equivalent to simply using the Mellin exchange amplitudes from Appendix \ref{App:B}, which have already been symmetrized (or anti-symmetrized), in (\ref{Stilde}). The difference is obvious in the MRV limit, as the symmetrized bosonic Mellin exchange amplitudes do not have improved u-channel Regge behavior. In principle, having specified the polar part of the amplitude there is still the possibility of adding contact terms. The truly distinguishing feature of our prescription, however, is that the {\it full} Mellin amplitude can be written as a sum of exchange amplitudes over multiplets, with {\it no} additional contact terms!\footnote{This in particular implies that there should exist a set of field redefinitions under which there are no quartic couplings for the scalars.} The Mellin amplitudes are just given by 
\begin{equation}
\mathcal{M}(s,t;\sigma,\tau)= \mathcal{M}_s(s,t;\sigma,\tau)+\mathcal{M}_t(s,t;\sigma,\tau)+\mathcal{M}_u(s,t;\sigma,\tau)\;,
\end{equation}
\begin{equation}
\mathcal{M}_s=\sum_p \mathcal{S}^{(s)}_p(s,t;\sigma,\tau)\;,\;\;
\mathcal{M}_t=\sum_p \mathcal{S}^{(t)}_p(s,t;\sigma,\tau)\;,\;\;
\mathcal{M}_u=\sum_p \mathcal{S}^{(u)}_p(s,t;\sigma,\tau)\;,
\end{equation}
with the multiplet amplitudes $\mathcal{S}^{(s)}_p$, $\mathcal{S}^{(t)}_p$, $\mathcal{S}^{(u)}_p$ obtained with the above prescription. The absence of the contact terms can be proven by the superconformal Ward identites, as we will discuss in detail in 
 Section \ref{Sec:5}.
 
Let us now rewrite the Mellin amplitude $\mathcal{M}(s,t;\sigma,\tau)$ into a different form that is more suitable for presentation. As we have seen, the Mellin amplitude has a series of simple poles at $s=\epsilon p_s+2m$, $t=\epsilon p_t+2m$, $u=\epsilon p_u+2m$, with
\begin{eqnarray}
p_s- \max\{|k_1-k_2|,|k_3-k_4|\}&=&2\;,4\;,\ldots\;,2\mathcal{E}-2\;,\\
p_t- \max\{|k_1-k_4|,|k_2-k_3|\}&=&2\;,4\;,\ldots\;,2\mathcal{E}-2\;,\\
p_u- \max\{|k_1-k_3|,|k_2-k_4|\}&=&2\;,4\;,\ldots\;,2\mathcal{E}-2\;.
\end{eqnarray}
A series of poles $s=\epsilon p_s+2m$ truncates  if
\begin{equation}
\epsilon(k_1+k_2)-\epsilon p_s= 2m_0\;,\;\; m_0\in \mathbb{Z}_+\;,\quad \text{or}\quad \epsilon(k_3+k_4)-\epsilon p_s= 2n_0\;,\;\; n_0\in \mathbb{Z}_+\;.
\end{equation}
The sum over $m$ is from 0 to $m_0-1$ or from 0 to $n_0-1$ if only one of them is an integer. In the case when both $m_0$ and $n_0$ are integers, $m$ is summed over from 0 to $\min\{m_0,n_0\}-1$. The truncation of poles in $t$ and $u$ is analogous. In the following we will write $\mathcal{M}_s(s,t;\sigma,\tau)$ as a sum over poles, and we decompose the numerators into different R-symmetry structures spanned by the monomials of $\sigma$, $\tau$
\begin{equation}
\mathcal{M}_s(s,t;\sigma,\tau)=\sum_{i,j}\sigma^i\tau^j\sum_{s_0}\frac{R^{i,j}_s(t,u)}{s-s_0}\;.
\end{equation}
The residues $R^{i,j}_{s_0}(t,u)$ are a sum over  supergravity multiplets labelled by the Kaluza-Klein level $p$ in the finite set (\ref{prange})
\begin{equation}
R^{i,j}_{s_0}(t,u)=\sum_p \mathcal{R}^{i,j}_{p,m}(t,u)\;, \quad \epsilon p+2 m=s_0\;,\;\;  m\in\mathbb{N}\;.
\end{equation}
The other two channels $\mathcal{M}_t(s,t;\sigma,\tau)$ and  $\mathcal{M}_u(s,t;\sigma,\tau)$ are similar, and can be obtained from  $\mathcal{M}_s(s,t;\sigma,\tau)$ by Bose symmetry. Using our method described above, we have calculated $\mathcal{R}^{i,j}_{p,m}(t,u)$ for all correlators in $AdS_4\times S^7$, $AdS_5\times S^5$ and $AdS_7\times S^4$. We will present their explicit expressions in the next subsection.

\subsection{All Mellin amplitudes for all maximally supersymmetric CFTs}
Let us define a set of convenient combinations $u^\pm$, $t^\pm$\begin{equation}
u^\pm=u\pm \frac{\epsilon}{2}\kappa_u-\frac{\epsilon}{2}\Sigma\;,\quad\;\; t^\pm=t\pm \frac{\epsilon}{2}\kappa_t-\frac{\epsilon}{2}\Sigma\;
\end{equation}
where we recall that $\epsilon=\frac{d-2}{2}$. We find that the residues from each multiplet take the universal form of 
\begin{equation}\label{residueR}
\mathcal{R}^{i,j}_{p,m}(t,u)= K^{i,j}_{p}(t,u)\, L^{i,j}_{p,m}\, N^{i,j}_{p}\;,
\end{equation}
in any spacetime dimension, and we give below the expressions for $ K^{i,j}_{p}$, $L^{i,j}_{p,m}$, $N^{i,j}_{p}$ in each background.

\vspace{0.5cm}
\noindent \underline{$AdS_5\times S^5$:}
\vspace{0.5cm}

\noindent Let us begin with the case of $d=4$, where the bulk theory is IIB supergravity on $AdS_5\times S^5$. The above procedure gives the following result 
\begin{eqnarray}
\nonumber K^{i,j}_{p}&=& 2i(2i+\kappa_u)t^-t^++2j(2j+\kappa_t)u^-u^+-2j\kappa_ut^+u^--2i\kappa_tu^+t^-\\
\nonumber &+& \frac{1}{4}(2p-\kappa_t-\kappa_u)(2p+\kappa_t+\kappa_u)(u^- t^-+4ij)\\
\nonumber &+&\frac{1}{2}(\kappa_u+\kappa_t-2p)(\kappa_u+\kappa_t+2p)(i t^-+ju^-)\\
 &+&4ij(t^+\kappa_u+u^+\kappa_t)-8ijt^+u^+\;,
\end{eqnarray}
\begin{equation}
L^{i,j}_{p,m}=\frac{(-1)^{i+j+\frac{2p-\kappa_t-\kappa_u}{4}}\prod_{i=1}^4\sqrt{k_i}}{n^2\,i!\,j!\,m!\,\Gamma[p+m+1]\Gamma[\tfrac{k_1+k_2-2m-p}{2}]\Gamma[\tfrac{k_3+k_4-2m-p}{2}]}\;,
\end{equation}
and
\begin{equation}
 N^{i,j}_{p}=\frac{2^{-3}p\,\Gamma[\frac{2p+\Sigma-\kappa_s-4l}{4}]}{\Gamma[\frac{\kappa_u+2+2i}{2}]\Gamma[\frac{2(p+2)-\Sigma+\kappa_s+4l}{4}]\Gamma[\frac{\kappa_t+2+2j}{2}]}
\end{equation}
where $i+j+l=\mathcal{E}$. Note that $L^{i,j}_{p,m}$ contains two Gamma factors $\Gamma[\tfrac{k_1+k_2-2m-p}{2}]\Gamma[\tfrac{k_3+k_4-2m-p}{2}]$ in the denominator. Since $k_i+k_j-p\in 2\mathbb{Z}_+$ by cubic vertex selection rules, they implement the truncation of poles in the Mellin amplitude.

 All tree-level four-point functions for $AdS_5\times S^5$ were given in \cite{Rastelli:2016nze,Rastelli:2017udc} after solving the bootstrap problem, and were written in terms of the reduced Mellin amplitude. The full amplitude can be obtained by acting with the superconformal difference operator $\widehat{R}$ (see \cite{Rastelli:2016nze,Rastelli:2017udc} for details). Upon comparing the residues, we find that above expressions reproduce the known result.

\vspace{0.5cm}
\noindent \underline{$AdS_7\times S^4$:}
\vspace{0.5cm}

\noindent  Next we turn to $d=6$, which corresponds to 11D supergravity on $AdS_7\times S^4$. The full solution to all four-point functions was recently obtained in \cite{Alday:2020lbp}. The residue factors are given by
\begin{eqnarray}
\nonumber K^{i,j}_{p}&=& 2i(2i+\kappa_u)t^-t^++2j(2j+\kappa_t)u^-u^++2j(1-\kappa_u)t^+u^-+2i(1-\kappa_t)u^+t^-\\
\nonumber &+& \frac{1}{4}(2p-\kappa_t-\kappa_u)(2p-2+\kappa_t+\kappa_u)(u^- t^-+16ij)\\
\nonumber &+&(\kappa_u+\kappa_t-2p)(\kappa_u+\kappa_t+2p-2)(i t^-+ju^-)\\
 &+&8ij(t^+(\kappa_u-1)+u^+(\kappa_t-1))-8ijt^+u^+\;,
\end{eqnarray} 
\begin{equation}
L^{i,j}_{p,m}=\frac{(-1)^{i+j+\frac{2p-\kappa_t-\kappa_u}{4}}\pi^{-\frac{3}{2}}\Gamma[\frac{k_1+k_2-p+1}{2}]\Gamma[\frac{k_3+k_4-p+1}{2}]\Gamma[\frac{k_1+k_2+p}{2}]\Gamma[\frac{k_3+k_4+p}{2}]}{n^3m!\, i!\, j!\prod_{a=1}^4\sqrt{(2k_a-2)!}\,\Gamma[2p+m] \Gamma[k_1+k_2-m-p]\Gamma[k_3+k_4-m-p]}\;,
 \end{equation}
\begin{equation}
 N^{i,j}_{p}=\frac{2^{\Sigma-6} (2p-1)\Gamma[\frac{2(p-1)+\Sigma-\kappa_s-4l}{4}]}{\Gamma[\frac{\kappa_u+2+2i}{2}]\Gamma[\frac{2(p+2)-\Sigma+\kappa_s+4l}{4}]\Gamma[\frac{\kappa_t+2+2j}{2}]}\;.
\end{equation}
The Gamma functions $\Gamma[k_1+k_2-m-p]\Gamma[k_3+k_4-m-p]$ in $L^{i,j}_{p,m}$ also ensure that the number of poles in the $AdS_7\times S^4$ Mellin amplitudes is finite.

\vspace{0.5cm}
\noindent \underline{$AdS_4\times S^7$:}
\vspace{0.5cm}

\noindent  Finally, we consider $d=3$ and it corresponds to 11D supergravity on $AdS_4\times S^7$. The only correlator which has been obtained in the literature is the four-point function of the stress tensor multiplet \cite{Zhou:2017zaw}. Here we present new results, which generalize to four-point functions of arbitrary one-half BPS operators
\begin{eqnarray}
\nonumber K^{i,j}_{p}&=& 2i(2i+\kappa_u)t^-t^++2j(2j+\kappa_t)u^-u^+-2j(2+\kappa_u)t^+u^--2i(2+\kappa_t)u^+t^-\\
\nonumber &+& \frac{1}{4}(2p-\kappa_t-\kappa_u)(4+2p+\kappa_t+\kappa_u)(u^- t^-+ij)\\
\nonumber&+&\frac{1}{4}(\kappa_u+\kappa_t-2p)(\kappa_u+\kappa_t+2p+4)(i t^-+ju^-)\\
 &+&2ij(t^+(2+\kappa_u)+u^+(2+\kappa_t))-8ijt^+u^+
\end{eqnarray}
\begin{eqnarray}
\nonumber L^{i,j}_{p,m}&=&\frac{\sqrt{\pi}\prod_{i=1}^4\sqrt{(k_i+1)!}}{n^{\frac{3}{2}}\,i!\,j!\,m!\,\Gamma[\tfrac{p+2m+3}{2}]\Gamma[\tfrac{k_1+k_2-p+2}{4}]\Gamma[\tfrac{k_3+k_4-p+2}{4}]\Gamma[\tfrac{k_1+k_2+p+4}{4}]\Gamma[\tfrac{k_3+k_4+p+4}{4}]}\\
&\times&\frac{(-1)^{i+j+\frac{2p-\kappa_t-\kappa_u}{4}}}{\Gamma[\tfrac{k_1+k_2-4m-p}{4}]\Gamma[\tfrac{k_3+k_4-4m-p}{4}]}\;.
\end{eqnarray}
\begin{equation}
 N^{i,j}_{p}=\frac{2^{-\frac{11+\Sigma}{2}}(1+p)\,\Gamma[\frac{2(p+2)+\Sigma-\kappa_s-4l}{4}]}{\Gamma[\frac{\kappa_u+2+2i}{2}]\Gamma[\frac{2(p+2)-\Sigma+\kappa_s+4l}{4}]\Gamma[\frac{\kappa_t+2+2j}{2}]}\;.
\end{equation}
Unlike the previous two cases, the Gamma function factors $\Gamma[\tfrac{k_1+k_2-4m-p}{4}]\Gamma[\tfrac{k_3+k_4-4m-p}{4}]$ in $L^{i,j}_{p,m}$ do not guarantee that the Mellin amplitudes should have a finite number of poles. Upon setting $k_i=2$, we reproduce the result of \cite{Zhou:2017zaw}.

\vspace{0.8cm}

Clearly, the Mellin amplitude residues in the three maximally supersymmetric backgrounds are highly similar. In fact, we can accentuate their similarity by writing down a formula which interpolates the M-theory and string theory amplitudes. More precisely, we can modify $K^{i,j}_{p}$, $L^{i,j}_{p,m}$, $N^{i,j}_{p}$ by introducing $\epsilon$-dependence as follows
\begin{eqnarray}
\nonumber K^{i,j}_{p}&=& 2i(2i+\kappa_u)t^-t^++2j(2j+\kappa_t)u^-u^+-2j(\tfrac{2}{\epsilon}-2+\kappa_u)t^+u^--2i(\tfrac{2}{\epsilon}-2+\kappa_t)u^+t^-\\
\nonumber &+& \frac{1}{4}(2p-\kappa_t-\kappa_u)(2p+\tfrac{4}{\epsilon}-4+\kappa_t+\kappa_u)(u^- t^-+4\epsilon^2 ij)\\
\nonumber &+&\tfrac{\epsilon}{2}(\kappa_u+\kappa_t-2p)(\kappa_u+\kappa_t+2p+\tfrac{4}{\epsilon}-4)(i t^-+ju^-)\\
 &+&4\epsilon ij(t^+(\kappa_u+\tfrac{2}{\epsilon}-2)+u^+(\kappa_t+\tfrac{2}{\epsilon}-2))-8ijt^+u^+\;,
 \end{eqnarray}
 \begin{eqnarray}
\nonumber L^{i,j}_{p,m}&=&\frac{\pi^{-\frac{(\epsilon-1)(2\epsilon+5)}{6}}2^{\frac{2(\epsilon-1)(2\epsilon-1)}{3}}\prod_{i=1}^4\big(\sqrt{k_i+\frac{1}{\epsilon}-1}\Gamma[\frac{2}{3}((1+\epsilon)k_i+2-\epsilon)]^{\frac{1}{3}(\frac{1}{\epsilon}-\epsilon)}\big)}{n^{1+\epsilon}\,\Gamma[2-\epsilon+m+\epsilon p]}\\
&\times&\frac{\big(\Gamma[\frac{(1+\epsilon)(k_1+k_2+p)}{6}+\frac{2(2-\epsilon)}{3}]\Gamma[\frac{(1+\epsilon)(k_3+k_4+p)}{6}+\frac{2(2-\epsilon)}{3}]\big)^{-\frac{2}{3}(\frac{1}{\epsilon}-\epsilon)}}{i!\,j!\,m!}\\
\nonumber&\times&\frac{(-1)^{i+j+\frac{2p-\kappa_t-\kappa_u}{4}}\big(\Gamma[\frac{(1+\epsilon)(k_1+k_2-p)}{6}+\frac{1}{2}]\Gamma[\frac{(1+\epsilon)(k_3+k_4-p)}{6}+\frac{1}{2}]\big)^{-\frac{2}{3}(\frac{1}{\epsilon}-\epsilon)}}{\Gamma[\frac{\epsilon}{2}(k_1+k_2-p)-m]\Gamma[\frac{\epsilon}{2}(k_3+k_4-p)-m]}\;,
 \end{eqnarray}
 and
 \begin{eqnarray}
 N^{i,j}_{p}&=&\frac{2^{\Sigma(\epsilon-1)-4-\epsilon}\Gamma[\frac{1}{4}(\frac{4}{\epsilon}-4+2p+\Sigma-\kappa_s-4l)](-(\frac{5\epsilon^2-15\epsilon+6}{\epsilon})p+1-\epsilon)}{\Gamma[\frac{\kappa_u+2+2i}{2}]\Gamma[\frac{2(p+2)-\Sigma+\kappa_s+4l}{4}]\Gamma[\frac{\kappa_t+2+2j}{2}]}\;.
\end{eqnarray}
When substituting in $\epsilon=\frac{1}{2}\,,\,1\,,\,2$, the above formulae reduce to the results in respective dimensions. Of course, such interpolation formulae that go through the three physical $\epsilon$ values are far from being unique, and we do not expect on any grounds that M-theory and string theory correlators should be physically connected. Nevertheless, what we wish to highlight is the similarities of analytic structures in the residues, which allow them to be compactly encapsulated in a single set of formulae. We also want to mention that the above sum over the multiplets $p$ can be performed in a closed form, and leads to a hypergeometric series. However,  we think that it is better to leave the sum unperformed, which makes the analytic structure more clear.

\subsection{Examples}
\label{examples}
Let us demonstrate our general formulae with a few illuminating examples.   The simplest example has $k_i=2$, which corresponds to the stress tensor four-point functions. The extremality $\mathcal{E}$ is 2.   Therefore, $i$, $j$ run from 0 to 2, and the Mellin amplitudes are degree-2 polynomials in $\sigma$, $\tau$. There is only one value $p=2$ in the range of summation (\ref{prange}), which means only the stress tensor multiplet contributes. Using our formulae, we find that for $\epsilon=1$
\begin{eqnarray}
\nonumber \mathcal{M}_{2222}^{AdS_5}(s,t;\sigma,\tau)&=&\mathcal{M}_{2222,s}^{AdS_5}(s,t;\sigma,\tau)+\mathcal{M}_{2222,t}^{AdS_5}(s,t;\sigma,\tau)+\mathcal{M}_{2222,u}^{AdS_5}(s,t;\sigma,\tau)\;,\\
\mathcal{M}_{2222,s}^{AdS_5}(s,t;\sigma,\tau)&=&-\frac{2}{n^2}\bigg(\frac{(t-4)(u-4)+(s+2)((t-4)\sigma+(u-4)\tau)}{s-2}\bigg)\;,\\
\nonumber \mathcal{M}_{2222,t}^{AdS_5}(s,t;\sigma,\tau)&=& \tau^2\mathcal{M}_{2222,s}^{AdS_5}(t,s;\tfrac{\sigma}{\tau},\tfrac{1}{\tau})\,,\; \mathcal{M}_{2222,u}^{AdS_5}(s,t;\sigma,\tau)= \sigma^2\mathcal{M}_{2222,s}^{AdS_5}(u,t;\tfrac{1}{\sigma},\tfrac{\tau}{\sigma})
\end{eqnarray}
where $s+t+u=8$. For $\epsilon=2$, we get 
\begin{eqnarray}
\nonumber \mathcal{M}_{2222}^{AdS_7}(s,t;\sigma,\tau)&=&\mathcal{M}_{2222,s}^{AdS_7}(s,t;\sigma,\tau)+\mathcal{M}_{2222,t}^{AdS_7}(s,t;\sigma,\tau)+\mathcal{M}_{2222,u}^{AdS_7}(s,t;\sigma,\tau)\;,\\
\nonumber \mathcal{M}_{2222,s}^{AdS_7}(s,t;\sigma,\tau)&=&-\frac{1}{n^3}\bigg(\frac{(t-8)(u-8)+(s+2)((t-8)\sigma+(u-8)\tau)}{s-4}\\
&&+\frac{(t-8)(u-8)+(s+2)((t-8)\sigma+(u-8)\tau)}{4(s-6)}\bigg)\;,\\
\nonumber \mathcal{M}_{2222,t}^{AdS_7}(s,t;\sigma,\tau)&=& \tau^2\mathcal{M}_{2222,s}^{AdS_7}(t,s;\tfrac{\sigma}{\tau},\tfrac{1}{\tau})\,,\; \mathcal{M}_{2222,u}^{AdS_7}(s,t;\sigma,\tau)= \sigma^2\mathcal{M}_{2222,s}^{AdS_7}(u,t;\tfrac{1}{\sigma},\tfrac{\tau}{\sigma})
\end{eqnarray}
where $s+t+u=16$. These two correlators respectively reproduce the results of \cite{Arutyunov:2000py} and \cite{Arutyunov:2002ff}. When $\epsilon=\frac{1}{2}$, we have 
\begin{eqnarray}
\label{M2222}
\nonumber \mathcal{M}_{2222}^{AdS_4}(s,t;\sigma,\tau)&=&\mathcal{M}_{2222,s}^{AdS_4}(s,t;\sigma,\tau)+\mathcal{M}_{2222,t}^{AdS_4}(s,t;\sigma,\tau)+\mathcal{M}_{2222,u}^{AdS_4}(s,t;\sigma,\tau)\;,\\
\mathcal{M}_{2222,s}^{AdS_4}(s,t;\sigma,\tau)&=&\sum_{m=0}^{\infty}-\frac{3((t-2)(u-2)+(s+2)((t-2)\sigma+(u-2)\tau))}{\sqrt{2 \pi } n^{\frac{3}{2}} \Gamma \left(\frac{1}{2}-m\right)^2 m! \Gamma \left(m+\tfrac{5}{2}\right)(s-1-2 m)}\;,\\
\nonumber \mathcal{M}_{2222,t}^{AdS_4}(s,t;\sigma,\tau)&=& \tau^2\mathcal{M}_{2222,s}^{AdS_4}(t,s;\tfrac{\sigma}{\tau},\tfrac{1}{\tau})\,,\; \mathcal{M}_{2222,u}^{AdS_4}(s,t;\sigma,\tau)= \sigma^2\mathcal{M}_{2222,s}^{AdS_4}(u,t;\tfrac{1}{\sigma},\tfrac{\tau}{\sigma})
\end{eqnarray}
where $s+t+u=4$. This reproduces the result of \cite{Zhou:2017zaw}, where the contact terms have now been automatically absorbed in the exchange contribution by our prescription.

Another interesting case is the next-to-next-to-extremal correlators with $k_1=k_2=2$, $k_3=k_4=k$. Let us only give the explicit result for $AdS_4\times S^7$,  which has not appeared in the literature. This family of correlators will be the starting point for constructing the four-point function $\langle2222\rangle$ at one loop. These correlators also have $\mathcal{E}=2$. Therefore, $p=2$ for the s-channel exchanges while $p=k$ for the t- and u-channel exchanges. We have 
\begin{eqnarray}
&&\mathcal{M}_{22kk,s}^{AdS_4}(s,t;\sigma,\tau)=\sum_{m=0}^\infty -\frac{3k}{8 \sqrt{2 \pi } n^{\frac{3}{2}} m! \Gamma [\frac{k-2 m-1}{2}]\Gamma[\frac{1-2m}{2}]\Gamma[\frac{5+2m}{2}]}\\
\nonumber &&\quad\quad\quad\quad\times \frac{(2 t-k-2) (2 u-k-2)+4 (s+2) \left(\sigma  \left(t-\tfrac{k}{2}-1\right)+\tau  \left(u-\tfrac{k}{2}-1\right)\right)}{s-1-2m}\;,
\end{eqnarray}
where $s+t+u=2+k$, and
\begin{eqnarray}
&&\mathcal{M}_{22kk,t}^{AdS_4}(s,t;\sigma,\tau)=\sum_{m=0}^\infty-\frac{3 k \tau  \Gamma[\frac{k}{2}+1]}{8 \sqrt{2} n^{\frac{3}{2}}  m!\Gamma [\frac{k-1}{2}] \Gamma[\frac{1-2m}{2}]^2  \Gamma[\frac{k+3+2m}{2}]}\\
\nonumber&&\quad\quad\quad\quad\times \frac{(2t+k+2) (2u-k-2)+2 (s-k) (\sigma  (k+2 t+2)+\tau  (2 u-k-2))}{t-\frac{k}{2}-2m}\;,\\
&&\mathcal{M}_{22kk,u}^{AdS_4}(s,t;\sigma,\tau)= \mathcal{M}_{22kk,t}^{AdS_4}(s,u;\tau,\sigma)\;.
\end{eqnarray}
Note that when $k$ is odd the pole series in $\mathcal{M}_{22kk,s}^{AdS_4}$ truncates, while if $k$ is even this does not happen.

Finally, let us give an example with higher extremality $\mathcal{E}=3$. We will consider the case with $k_i=3$. In the sum over multiplets, $p$ now takes values 2 and $4$ according to (\ref{prange}). Using our formulae, we get 
\begin{eqnarray}
&&\mathcal{M}_{3333,s}^{AdS_4}(s,t;\sigma,\tau)=\sum_{m=0}^\infty-\frac{27((t-3)(u-3)+(s+2)((t-3)\sigma+(u-3)\tau))}{4n^{\frac{3}{2}}\sqrt{2\pi}m!\Gamma[1-m]^2\Gamma[\frac{2m+5}{2}](s-1-2m)}\\
\nonumber&&\quad\quad\quad\quad+\sum_{m=0}^\infty\frac{48}{5n^{\frac{3}{2}}\sqrt{2\pi}m!\Gamma[\frac{1-2m}{2}]^2\Gamma[\frac{2m+7}{2}](s-2-2m)}\\
\nonumber&&\quad\quad\quad\quad\quad\times\big[(t-3)(u-3)+4(s+3)((s+2)\sigma\tau-(t-3)\sigma^2-(u-3)\tau^2)\\
\nonumber &&\quad\quad\quad\quad\quad+(s+3)((t-3)\sigma+(u-3)\tau)-4((t-3)(u-\tfrac{15}{4})\sigma+(u-3)(t-\tfrac{15}{4})\tau)\big]
\end{eqnarray}
where $s+t+u=6$. The other two channels are related by crossing symmetry
\begin{eqnarray}
\nonumber\mathcal{M}_{3333,t}^{AdS_4}(s,t;\sigma,\tau)&=& \tau^3\mathcal{M}_{3333,s}^{AdS_4}(t,s;\tfrac{\sigma}{\tau},\tfrac{1}{\tau})\,,\\
\mathcal{M}_{3333,u}^{AdS_4}(s,t;\sigma,\tau)&=& \sigma^3\mathcal{M}_{3333,s}^{AdS_4}(u,t;\tfrac{1}{\sigma},\tfrac{\tau}{\sigma})\;.
\end{eqnarray}

\section{Superconformal Ward identities}\label{Sec:5}
\subsection{WI in Mellin space}
In the previous section we have constructed the polar part of the general Mellin amplitudes for the backgrounds $AdS_4 \times S^7$, $AdS_5 \times S^5$ and $AdS_7 \times S^4$, and claimed that no further contact terms are needed. In order to show that these contact terms are absent, we need to show that these amplitudes satisfy the superconformal Ward Identities (WI). Note that since the WI were not heavily used in our construction, this also serves as a non-trivial check of our results. In the cases of $AdS_5 \times S^5$ and $AdS_7 \times S^4$ one can efficiently impose the WI by requiring the existence of a reduced amplitude $\widetilde{\cal M}$, as discussed in Section \ref{Sec:2.2}. However, for  $AdS_4 \times S^7$ this is not possible. Below we will develop an efficient method to impose the WI in Mellin space at the level of the full amplitude, expanding on \cite{Zhou:2017zaw}. We start by recalling the WI (\ref{scfWardid}) in space-time
\begin{equation}\label{WIst}
(z\partial_z-\epsilon \alpha\partial_\alpha)\mathcal{G}(z,\bar{z};\alpha,\bar{\alpha})\big|_{\alpha=1/z}=0\;. 
\end{equation}
In order to write this relation in Mellin space we first note
\begin{equation}
z \partial_z = U \partial_U - \frac{z}{1-z}V \partial_V
\end{equation}
In Mellin space $U \partial_U$ and $V \partial_V$ have a very simple, multiplicative, action, which follows from the definition (\ref{inverseMellin})
\begin{equation}
U \partial_U \to \left(\frac{s}{2}-a_s \right) \times,~~~V \partial_V \to \left(\frac{t}{2}-a_t \right) \times
\end{equation}
On the other hand, $z$ does not. In order to proceed we write the Mellin amplitude in terms of the R-symmetry cross ratios $\alpha,\bar \alpha$ and expand it in powers of $\alpha$:
\begin{equation}
{\cal M}(s,t,\alpha,\bar\alpha)= \sum_{q=0}^{\mathcal{E}}  \alpha^q {\cal M}^{(q)}(s,t,\bar \alpha)
\end{equation}
In terms of the components ${\cal M}^{(q)}(s,t,\bar \alpha)$ the WI take the form
\begin{equation}
\label{WIMellin}
\sum_{q=0}^{\mathcal{E}} \left( (1-z)z^{\mathcal{E}-q} \left(\frac{s}{2}-a_s -q\right) -z^{\mathcal{E}-q+1} \left(\frac{t}{2}-a_t \right) \right){\cal M}^{(q)}(s,t,\bar \alpha) = 0.
\end{equation}
We can obtain an inequivalent relation by replacing $z \to \bar z$. 
\begin{equation}
\label{WIMellinc}
\sum_{q=0}^{\mathcal{E}} \left( (1-\bar z)\bar z^{\mathcal{E}-q} \left(\frac{s}{2}-a_s -q\right) -\bar z^{\mathcal{E}-q+1} \left(\frac{t}{2}-a_t \right) \right){\cal M}^{(q)}(s,t,\bar \alpha) = 0.
\end{equation}
Considering two independent linear combinations of the relations above we arrive at
\begin{equation}
\label{WIMellinzeta}
\sum_{q=0}^{\mathcal{E}} \left( (\zeta_{\pm}^{(\mathcal{E}-q)}- \zeta_{\pm}^{(\mathcal{E}-q+1)}) \left(\frac{s}{2}-a_s -q\right) -\zeta_{\pm}^{(\mathcal{E}-q+1)} \left(\frac{t}{2}-a_t \right) \right){\cal M}^{(q)}(s,t,\bar \alpha) = 0.
\end{equation}
where we have defined
\begin{equation}
\zeta_+^{(n)} = z^n + \bar z^n,~~~~\zeta_-^{(n)} = \frac{z^n - \bar z^n}{z-\bar z}
\end{equation}
The crucial observation is that, while $z$ and $\bar z$ by themselves do not have a simple action in Mellin space, $\zeta_{\pm}^{(n)}$, which should be interpreted as operators, do. Indeed, for each $n$ $\zeta_{\pm}^{(n)}$ are simply polynomials of $U$ and $V$, while powers of $U$ and $V$ act in Mellin space as shift operators. This leads to the following representation in Mellin space
\begin{align}
&\zeta_+^{(0)} = 2,~~~~&\zeta_-^{(0)}= 0\\
 &\zeta_+^{(1)} = 1+\widehat{U}-\widehat{V},~~~~&\zeta_-^{(1)} = 1\\
  &\zeta_+^{(2)} = 1-2 \widehat{V}+\widehat{U^2}+\widehat{V^2}- 2 \widehat{U V},~~~~&\zeta_-^{(2)} = 1+\widehat{U}-\widehat{V}
\end{align}
and so on, where $\widehat{U^m V^n}$ is the shift operator corresponding to $U^m V^n$ and is given by 
\begin{equation}
\widehat{U^m V^n} \circ {\cal M}(s,t) = \frac{\Gamma_{\{k_i\}}(s-2m,t-2n)}{\Gamma_{\{k_i\}}(s,t)}{\cal M}(s-2m,t-2n).
\end{equation}
Note that for a given extremality ${\mathcal{E}}$ only operators up to $\zeta_{\pm}^{({\mathcal{E}}+1)}$ appear. 

\subsubsection*{An example}
The simplest example is that of $k_i=2$, namely the correlator of the stress-tensor multiplet. So let us work out this case in detail. We will focus in the equation (\ref{WIMellinzeta}) involving $\zeta_-$, which has not been explicitly considered before. In this case the extremality ${\mathcal{E}}=2$ and we can decompose the Mellin amplitude as
\begin{equation}
{\cal M}(s,t,\alpha,\bar\alpha)={\cal M}^{(0)}(s,t)+ \alpha {\cal M}^{(1)}(s,t)+\alpha^2 {\cal M}^{(2)}(s,t)
\end{equation}
where the dependence on $\bar \alpha$ has not been explicitly shown, since it acts as a spectator. The WI takes the form
\begin{eqnarray}
 &&\zeta_-^{(1)} \left( (s+t-8\epsilon) {\cal M}^{(2)}(s,t)+(2\epsilon-s){\cal M}^{(1)}(s,t) \right) \\ 
\nonumber &&+\zeta_-^{(2)} \left( (s+t-6\epsilon) {\cal M}^{(1)}(s,t)-s{\cal M}^{(0)}(s,t) \right) + \zeta_-^{(3)} (s+t-4\epsilon) {\cal M}^{(0)}(s,t) = 0\;,
\end{eqnarray}
or explicitly after acting with the shift operators
\begin{align}
\label{WIE2}
&(t-4\epsilon){\cal M}^{(0)}(s,t) -\frac{2 (s-4 \epsilon )^2 (t-4 \epsilon )^2}{(s+t-4 \epsilon -2)^2 (s+t-4 (\epsilon +1))}{\cal M}^{(0)}(s-2,t-2) \nonumber \\
&+ \frac{\left(s^2-2 s (4 \epsilon +1)+8 \epsilon  (2 \epsilon +1)\right)^2}{(s+t-4 \epsilon -2)^2 (s+t-4 (\epsilon +1))} {\cal M}^{(0)}(s-4,t) -\frac{(t-4 \epsilon )^2 (s+2 t-8 \epsilon -4)}{(s+t-4 \epsilon -2)^2} {\cal M}^{(0)}(s,t-2)  \nonumber\\
&+ \frac{\left(t^2-2 t (4 \epsilon +1)+8 \epsilon  (2 \epsilon +1)\right)^2}{(s+t-4 \epsilon -2)^2 (s+t-4 (\epsilon +1))} {\cal M}^{(0)}(s,t-4) +\frac{(s-4 \epsilon )^2 (t-4 \epsilon )}{(s+t-4 \epsilon -2)^2} {\cal M}^{(0)}(s-2,t)  \nonumber\\
&+ \frac{(s-4 \epsilon )^2 (s+t-6 \epsilon -2)}{(s+t-4 \epsilon -2)^2} {\cal M}^{(1)}(s-2,t) -\frac{(t-4 \epsilon )^2 (s+t-6 \epsilon -2)}{(s+t-4 \epsilon -2)^2} {\cal M}^{(1)}(s,t-2)  \nonumber \\
&+(t-4\epsilon){\cal M}^{(1)}(s,t)+(s+t-8\epsilon) {\cal M}^{(2)}(s,t) = 0\;.
\end{align}
Note that this gives ${\cal M}^{(2)}(s,t)$ in terms of ${\cal M}^{(0)}(s,t)$ and ${\cal M}^{(1)}(s,t)$. This is a general phenomenon: For a general extremality ${\mathcal{E}}$, we can use the WI involving $\zeta_-$ to solve for ${\cal M}^{({\mathcal{E}})}(s,t)$ in terms of the other ones. Returning to (\ref{WIE2}), for $d=4,6$ we can simply plug the results given in Section \ref{examples} and check that they indeed satisfy this relation for $\epsilon=1$ and $\epsilon=2$ respectively. For $d=3$ we can resum the expression given in (\ref{M2222}) to obtain
\begin{eqnarray}
\mathcal{M}_{2222,s}^{AdS_4}(s,t;\sigma,\tau)=& \left( -\frac{3 (t-2) (u-2)}{2 \sqrt{2} \pi ^{3/2} n^{3/2} (s-1) s (s+2) \Gamma [1-\frac{s}{2}]} + \frac{3 \sqrt{2} (t-2) (t+u-6)}{\pi ^{3/2} n^{3/2} (s-1) s^2 (s+2)^2 \Gamma [-\frac{s}{2}-1]}\sigma \nonumber \right.\\
&\left. +\frac{3 \sqrt{2} (u-2) (t+u-6)}{\pi ^{3/2} n^{3/2} (s-1) s^2 (s+2)^2 \Gamma [-\frac{s}{2}-1]} \tau \right) h(s)
\end{eqnarray}
where we have introduced 
\begin{equation}
h(s) = \sqrt{\pi } \left(s^2+3 s-4\right) \Gamma \left[1-\frac{s}{2}\right]+8 \Gamma \left[\frac{3-s}{2}\right].
\end{equation}
Adding the contributions in the t- and u-channels we can obtain the corresponding expressions for ${\cal M}^{(q)}(s,t)$, for $q=0,1,2$. Plugging them into (\ref{WIE2}) we can check that indeed, the identity is satisfied for $\epsilon=1/2$.

We have checked the above WI for a vast variety of examples. We have found our answer satisfies the WI in each case, without the addition of a contact term. This actually proves that by using the representation we have chosen, our results provides the full answer and not just the polar part of the amplitude.  

\subsection{WI and the flat space limit}
It is illuminating to study the superconformal Ward identities and the Mellin amplitudes around the flat space limit, where $s,t$ are large. In the flat space limit shift operators act multiplicatively. Indeed in this limit ${\cal M}(s-2m,t-2n) \sim {\cal M}(s,t)$ plus higher order derivative corrections, and one can explicitly check
\begin{equation}
\widehat{U^m V^n} \circ {\cal M}(s,t) = \frac{s^{2m}t^{2n}}{(s+t)^{2(m+n)}} + \cdots.
\end{equation}
This leads to the following rule for the operators $\zeta_{\pm}^{(n)}$ to leading order
\begin{equation}
\zeta_{+}^{(n)} = \frac{2s^n}{(s+t)^n}+ \cdots,~~~~\zeta_{-}^{(n)} = \frac{n s^{n-1}}{(s+t)^{n-1}}+ \cdots.
\end{equation}
Plugging these expressions in (\ref{WIMellinzeta}) and taking the flat space limit, we observe the equation for $\zeta_{+}$ is trivially satisfied to leading order, while the remaining equation gives
\begin{equation}
\sum_{q=0}^{\mathcal{E}} \frac{s^{\mathcal{E}-q}}{(s+t)^{\mathcal{E}-q}} {\cal M}^{(q)}_{\text{flat}}(s,t,\bar \alpha) = 0.
\end{equation}
But this simply implies that in the flat space limit
\begin{eqnarray}
\label{WIflat}
{\cal M}_{\text{flat}}(s,t; \frac{s+t}{s}, \bar \alpha) = 0,\\
{\cal M}_{\text{flat}}(s,t; \alpha,\frac{s+t}{s}) = 0. \nonumber
\end{eqnarray}
as a consequence of the superconformal Ward identities, in any number of dimensions. The second relation follows from replacing $\alpha \to \bar \alpha$.

From our results, we can study the explicit form of the amplitudes in the flat space limit. In all cases we find
\begin{equation}
\nonumber
\lim_{s,t\to \infty}{\cal M}(s,t;\sigma,\tau) = {\cal N}_{\{k_i\}} \frac{\Theta_4^{\text{flat}}(s,t;\sigma,\tau)}{s t u} P_{\{k_i\}}(\sigma,\tau)
\end{equation}   
with $s+t+u=0$ in the flat space limit, and 
\begin{equation}
\Theta_4^{\text{flat}}(s,t;\sigma,\tau) = \left(t u + t s \sigma + s u \tau \right)^2.
\end{equation}
$P_{\{k_i\}}(\sigma,\tau)$ is an R-symmetry polynomial explicitly given by
\begin{equation}
\nonumber P_{\{k_i\}} = \sum_{\substack{i+j+k = \mathcal{E} -2 \\ 0 \leq i,j,k \leq \mathcal{E}-2}} \frac{(\mathcal{E}-2)!\, \sigma^i \tau^j}{i!\,j!\,k!\, (i+\tfrac{\kappa_u}{2})!\, (j+\tfrac{\kappa_t}{2})!\, (k+\tfrac{\kappa_s}{2})!}\;.
\end{equation}
Note that the form of the flat-space limit is completely universal, and the prefactor $\Theta_4^{\text{flat}}$ as well as the polynomials $P_{\{k_i\}}(\sigma,\tau)$ do not depend on the number of dimensions. Furthermore, rewriting $\Theta_4^{\text{flat}}$ in terms of $\alpha,\bar \alpha$ and using $s+t+u=0$ we obtain 
\begin{equation}
\Theta_4^{\text{flat}}(s,t;\alpha,\bar \alpha) = (s+t-s \alpha)^2(s+t-s \bar \alpha)^2
\end{equation}
which neatly factorizes into a holomorphic and an anti-holomorphic part. Note that the presence of this factor implies the relations (\ref{WIflat}) indeed hold. For $d=4,6$ the presence of the prefactor $\Theta_4^{\text{flat}}(s,t;\alpha,\bar \alpha)$ in the flat space limit has also been discussed in  \cite{Chester:2018aca,Chester:2018dga}. In those cases the solutions to the WI can be written as a shift operator acting on a reduced amplitude, and we can show that the flat space limit of such shift operator always contains the prefactor $\Theta_4^{\text{flat}}(s,t;\alpha,\bar \alpha)$. 

\section{Conclusion}\label{Sec:6}
In this paper we developed a constructive method to obtain tree-level four-point holographic correlators in all theories with maximal superconformal symmetry. Our method exploits the remarkable simplicity of the Mellin amplitude at the MRV limit, which hides new powerful organizing principles for holographic correlators. The construction of the full amplitude from this limit is universal for all spacetime  dimensions, and allows us to derive results for different backgrounds on the same footing. For $d=4$, our result constitutes a proof for a widely believed conjecture \cite{Rastelli:2016nze,Rastelli:2017udc}. For $d=6$, we reproduce the results recently reported in \cite{Alday:2020lbp}, and for $d=3$ we provide new results. Our results lead to an array of interesting questions, applications, and  avenues for future research. We list a few below.
\begin{itemize}
\item The four-point functions we have constructed contain a wealth of CFT data. For $d=3$, part of these data can be compared with other exact results from topological twisting and supersymmetric localization \cite{Chester:2014fya,Chester:2014mea,Beem:2016cbd,Mezei:2017kmw}. They can also be used to calibrate the numerical bootstrap bounds at large central charge \cite{Chester:2014fya,Chester:2014mea,Agmon:2017xes}.
\item What we have done in this paper can also be viewed as the first step towards carrying out the program of computing loops in maximally supersymmetric supergravity theories, where the tree-level correlators gives essential input for applying the AdS unitarity method \cite{Aharony:2016dwx}. While this program is quite advanced  in $AdS_5 \times S^5$ \cite{Alday:2017xua,Aprile:2017bgs,Aprile:2017xsp,Alday:2017vkk,Aprile:2017qoy,Aprile:2018efk,Caron-Huot:2018kta,Alday:2018pdi,Alday:2018kkw,Aprile:2019rep,Alday:2019nin,Bissi:2020wtv}, it is still in its infancy for $AdS_7 \times S^4$ \cite{Alday:2020tgi}.   Similar progress  for $AdS_4 \times S^7$ at one loop yet awaits being made.

\item In our construction we give a prescription for restoring Bose symmetry in the exchange amplitudes, which at the same time allows the full amplitude to be expressed as a sum over exchange amplitudes with no extra contact terms. The absence of contact terms is a clear indication of on-shell reconstructibility in AdS, and a similar phenomenon was also observed at the level of the five-point function \cite{Goncalves:2019znr}. It would be interesting to have a better understanding of the observed reconstructibility, which could be useful for finding efficient algorithms to construct higher-point correlators.  

\item It would be very interesting to generalize what we have done to non-maximally supersymmetric CFTs in $d>2$. Some initial progress using bootstrap methods has been reported in \cite{Zhou:2018ofp} for four-point functions of lowest KK modes. We expect that using the MRV limit will fix the contributions from within each multiplet more efficiently than imposing the Mellin superconformal Ward identities, and therefore streamlines the calculation for higher KK modes. Moreover, it would be interesting to see if the same prescription will continue to absorb the contact terms into the exchange amplitudes when there is less supersymmetry present.  

\item We can also study various other limits of the general four-point correlators. One interesting limit is to take $k_i$ large, where we would expect to see the semiclassical behavior of membranes or strings scattering in AdS. 
\item We have also initiated a study of the Mellin superconformal WI (and their solutions) around the flat space limit. It may be interesting to pursue this further to construct the solution to the WI for the $d=3$ case, where the solution in position space contains non-local differential operators.
\item There has been some progress in understanding gravitational MHV amplitudes through twistor actions in the presence of a cosmological constant (see \cite{Adamo:2013tja} and references therein). It would be very interesting to make a connection between that formalism and the results of this paper.
\end{itemize}

\acknowledgments
The work of LFA is supported by the
European Research Council (ERC) under the European Union's Horizon 2020 research and
innovation programme (grant agreement No 787185). The work of XZ is supported in part by the Simons Foundation Grant No. 488653.

\appendix

\section{R-symmetry polynomials}\label{App:A}
In this appendix we present the R-symmetry polynomials $Y_{mn}^{(a,b)}(\sigma,\tau)$ for the compact group $SO(\mathtt{d})$. We follow the notation in \cite{Nirschl:2004pa}. In terms of these, the ones used in the body of the paper are given by
\begin{equation}
{\cal Y}_{\{d_1,d_2\}} = Y^{(\frac{\kappa_t}{2},\frac{\kappa_u}{2})}_{\frac{d_1+d_2}{2}-\frac{\kappa_t+\kappa_u}{4},\frac{d_1}{2}-\frac{\kappa_t+\kappa_u}{4}}.
\end{equation}
$Y_{mn}^{(a,b)}(\sigma,\tau)$ are eigenfunctions of the $SO(\mathtt{d})$ Casimir operator
\begin{equation}
\label{Casimir}
L^2 Y^{(a,b)}_{mn}(\sigma,\tau) = -2 C_{mn}^{(a,b)} Y^{(a,b)}_{mn}(\sigma,\tau) 
\end{equation}
where we define 
\begin{eqnarray}
\nonumber L^2 &=&2 {\cal D}_\mathtt{d}^{(a,b)} -\tfrac{1}{2}(a+b)(a+b+\mathtt{d}-2)\,,\\
C_{mn}^{(a,b)} &=& (m+\tfrac{a+b}{2})(m+\tfrac{a+b}{2}+\mathtt{d}-3)+(n+\tfrac{a+b}{2})(n+\tfrac{a+b}{2}+1)\,, \nonumber
\end{eqnarray}
with ${\cal D}_\mathtt{d}^{(a,b)}$ given by
\begin{eqnarray}
 {\cal D}_\mathtt{d}^{(a,b)}&=& {\cal D}_\mathtt{d}+(1-\sigma-\tau)\left( a \partial_\tau + b \partial_\sigma \right) -2 a \sigma \partial_\sigma-2b \tau \partial_\tau,\\
 {\cal D}_\mathtt{d} &=& (1-\sigma-\tau) \left( \partial_\sigma \sigma \partial_\sigma + \partial_\tau \tau \partial_\tau \right) - 4 \sigma \tau \partial_\sigma \partial_\tau - (\mathtt{d} -2) \left( \sigma \partial_\sigma + \tau \partial_\tau \right).
\end{eqnarray}
The R-symmetry polynomials admit the following expansion
\begin{eqnarray}
Y^{(a,b)}_{mn}(\sigma,\tau) = \sum_{i,j=0}^{i+j=m} c_{i,j} \sigma^i \tau^j\,,
\end{eqnarray}
where the coefficients $c_{i,j}$ take the form
\begin{equation}
\label{cansatz}
c_{i,j} = P^{(m-n)}(i,j)  \frac{(-1)^{m-i-j} \Gamma \left(a+b+\frac{\mathtt{d}}{2}+i+j+n-1\right)}{ \Gamma (a+j+1) \Gamma (b+i+1)\Gamma (i+1) \Gamma (j+1) \Gamma (m-i-j+1)}  
\end{equation}
with $P^{(m-n)}(i,j)$ a polynomial in $i,j$ of total degree $m-n$. Recall $m \geq n$. We fix the overall normalization such that the highest power of $\sigma$ has coefficient $1$, namely $Y^{(a,b)}_{mn}(\sigma,\tau) = \sigma^m + \cdots$. In terms of the polynomial $P^{(m-n)}(i,j)$  this takes the form
\begin{equation}
P^{(m-n)}(m,0) =  \frac{\Gamma (a+1) \Gamma (m+1) \Gamma (m+b+1)}{\Gamma \left(a+b+\frac{\mathtt{d}}{2}+m+n-1\right)}\;.
\end{equation}
The Casimir equation is equivalent to the following recursion relation  for $P^{(m-n)}(i,j)$
\begin{eqnarray} 
\nonumber &&2 (i (2 a+b+\mathtt{d}+4 j-2)+j (a+2 b+\mathtt{d}-2)) P(i,j) +2 (i^2+j^2-n^2)P(i,j) \\
 &&-2 (m (a+b+\mathtt{d}+m-3)+n (a+b+1)) P(i,j)  \\
\nonumber&&  +2j(a+j) P(i+1,j-1) =(i+j-m) (\mathtt{d}+2 (a+b+i+j+n-1)) P(i+1,j) \\
\nonumber &&+ (i + j - m) (2 a + 2 b + \mathtt{d} + 2 (i + j + n-1)) P(i,j+1) - 2i(b+i) P(i-1,j+1)\;.
\end{eqnarray}
For any fixed degree $m-n$ the relation can be solved. Up to an overall factor the R-symmetry polynomials take the form
\begin{equation}
Y^{(a,b)}_{mn}= P^{(m-n)}(\sigma \partial_\sigma,\tau \partial_\tau) F_4\bigg[ \begin{matrix} -m & n+a+b+ \frac{\mathtt{d}}{2}-1 \\ b+1 & a+1 \end{matrix} ; \sigma, \tau\bigg]
\end{equation}
where we have introduced the Appell's generalized hypergeometric function $F_4$
\begin{equation}
F_4\bigg[ \begin{matrix} a & b \\ c & d \end{matrix} ; x, y\bigg] =\sum_{m,n} \frac{(a)_{m+n} (b)_{m+n}}{m! n! (c)_m (d)_n} x^m y^n.
\end{equation}

\section{Exchange Mellin amplitudes in $AdS_{d+1}$}\label{App:B}
In this appendix, we present expressions for exchange Mellin amplitudes in $AdS_{d+1}$ with generic conformal dimensions. We consider exchanged fields with dimension $\Delta_E$ and spin $\ell_E$ up to 2. The Mellin amplitudes take the form
\begin{equation}
\mathcal{M}_{\Delta_E,\ell_E}(s,t)=\sum_{m}\frac{f_{m,\ell_E}\, Q_{m,\ell_E}(t,u)}{s-\Delta_E+\ell_E-2m}\;,
\end{equation}
and the residues can be obtained by solving the Casimir equation in Mellin space. We have 
\begin{equation}
 f_{m,\ell_E}=\frac{(-1)\,2^{1-2\ell_E}\Gamma[\Delta_E+\ell_E]\big(\tfrac{2-\ell_E-\Delta^{1,2}_E}{2}\big)_m\big(\tfrac{2-\ell_E-\Delta^{3,4}_E}{2}\big)_m}{m!(\tfrac{2\Delta_E-d+2}{2})_m \Gamma[\tfrac{\Delta^{1,2}_E+\ell_E}{2}]\Gamma[\tfrac{\Delta^{3,4}_E+\ell_E}{2}]\Gamma[\tfrac{\Delta^{1,E}_2+\ell_E}{2}]\Gamma[\tfrac{\Delta^{2,E}_1+\ell_E}{2}]\Gamma[\tfrac{\Delta^{3,E}_4+\ell_E}{2}]\Gamma[\tfrac{\Delta^{4,E}_3+\ell_E}{2}]}
\end{equation}
where $\Delta^{i,j}_k\equiv \Delta_i+\Delta_j-\Delta_k$.
$Q_{m,\ell_E}(t,u)$ are polynomials in $t$ and $u$ of degree $\ell_E$, and are given by
\begin{eqnarray}
\nonumber Q_{m,0}&=&1\;,\\
\nonumber Q_{m,1}&=& \frac{(\delta_u^2-\delta_t^2)(t+u+d-2-\Sigma_\Delta)}{4(\Delta_E-d+1)}+(\Delta_E-1)(t-u)\;,\\
Q_{m,2}&=& \frac{(d-1)T_1}{16d(\Delta_E-d)}-\frac{(d-1)T_2}{16d(\Delta_E-d+1)}+\frac{T_3}{16d}\\
\nonumber &+&\frac{(\delta_u^2-\delta_t^2)}{2}(t-u)(u+t+d-2-\Sigma_\Delta)\\
\nonumber &-&\frac{2(1-\Delta_E+\Delta_E^2)-\delta_t^2-\delta_u^2}{2d}(u+t+d-2-\Sigma_\Delta)^2\\
\nonumber &-&\Delta_E(1-\Delta_E)(t-u)^2\;,
\end{eqnarray}
where 
\begin{eqnarray}
\nonumber T_1&=&(\delta_u^2-\delta_t^2)(t+u+d-2-\Sigma_\Delta)\big(u(\delta_u^2-\delta_t^2-8d)\\
\nonumber &+&t (\delta_u^2-\delta_t^2+8d)-(\delta_u^2-\delta_t^2)(\Sigma_\Delta-d+2)\big)\;,\\
T_2&=&((\delta_u-\delta_t)^2-4)((\delta_u+\delta_t)^2-4)(t+u+d-3-\Sigma_\Delta)\\
\nonumber &\times& (t+u+d-1-\Sigma_\Delta)\;,\\
\nonumber T_3&=&((\delta_u-\delta_t)^2-4)((\delta_u+\delta_t)^2-4)\\
\nonumber &+&8\Delta_E(\Delta_E-1)(2(\Delta_E^2-d(\Delta_E+3)+d^2+1)-\delta_t^2-\delta_u^2)\;.
\end{eqnarray}
In the above we have defined the shorthand notations
\begin{eqnarray}
\nonumber\delta_t&\equiv&\Delta_1+\Delta_4-\Delta_2-\Delta_3\;,\\\delta_u&\equiv&\Delta_2+\Delta_4-\Delta_1-\Delta_3\;,\\\nonumber\Sigma_\Delta&\equiv&\Delta_1+\Delta_2+\Delta_3+\Delta_4\;.
\end{eqnarray}

\bibliography{refallmaxsusy} 
\bibliographystyle{utphys}

\end{document}